\documentclass{article}

\usepackage{microtype}
\usepackage{graphicx}
\usepackage{subfigure}
\usepackage{booktabs} % for professional tables

\usepackage{comment}
\usepackage{wrapfig}
\usepackage{amsfonts}
\usepackage{hyperref}
\usepackage{enumitem}

\usepackage{amsmath}
\usepackage{bbm}
\usepackage{algorithm}
\usepackage{algorithmic}
\usepackage{booktabs}
\usepackage{comment}
\usepackage{wrapfig}
\newtheorem{defn}{Definition}[section]

\usepackage[accepted]{mlsys2020}

\mlsystitlerunning{SLIDE : In Defense of Smart Algorithms over Hardware Acceleration for Large-Scale Deep Learning Systems}

\begin{document}

\twocolumn[
\mlsystitle{SLIDE : In Defense of Smart Algorithms over Hardware Acceleration for Large-Scale Deep Learning Systems}

% unless you've provided the [accepted] option to the mlsys2020


% \mlsyssetsymbol{equal}{*}

\begin{mlsysauthorlist}
\mlsysauthor{Beidi Chen}{rice}
\mlsysauthor{Tharun Medini}{rice}
\mlsysauthor{James Farwell}{intel}
\mlsysauthor{Sameh Gobriel}{intel}
\mlsysauthor{Charlie Tai}{intel}
\mlsysauthor{Anshumali Shrivastava}{rice}
\end{mlsysauthorlist}

\mlsysaffiliation{rice}{Rice University}
\mlsysaffiliation{intel}{Intel Corporation}

\mlsyscorrespondingauthor{Beidi Chen}{beidi.chen@rice.edu}
% \mlsyscorrespondingauthor{Eee Pppp}{ep@eden.co.uk}

\mlsyskeywords{Machine Learning, MLSys}

\vskip 0.3in

\begin{abstract}
Deep Learning (DL) algorithms are the central focus of modern machine learning systems. As data volumes keep growing, it has become customary to train large neural networks with hundreds of millions of parameters to maintain enough capacity to memorize these volumes and obtain state-of-the-art accuracy. To get around the costly computations associated with large models and data, the community is increasingly investing in specialized hardware for model training. However, specialized hardware is expensive and hard to generalize to a multitude of tasks. The progress on the algorithmic front has failed to demonstrate a direct advantage over powerful hardware such as NVIDIA-V100 GPUs. This paper provides an exception. We propose SLIDE (\textbf{S}ub-\textbf{LI}near \textbf{D}eep learning \textbf{E}ngine) that uniquely blends smart randomized algorithms, with multi-core parallelism and workload optimization. Using just a CPU, SLIDE drastically reduces the computations during both training and inference outperforming an optimized implementation of Tensorflow (TF) on the best available GPU. Our evaluations on industry-scale recommendation datasets, with large fully connected architectures, show that training with SLIDE on a 44 core CPU is more than 3.5 times (1 hour vs. 3.5 hours) faster than the same network trained using TF on Tesla V100 at any given accuracy level. On the same CPU hardware, SLIDE is over 10x faster than TF. We provide codes and scripts for reproducibility.
\end{abstract}
]


\printAffiliationsAndNotice{}  % leave blank if no need to mention equal contribution
% \printAffiliationsAndNotice{\mlsysEqualContribution} % otherwise use the standard text.

\section{Introduction}\label{submission}
Deep Learning (DL) has become a topic of significant interest in the research community. The last few years have seen a remarkable growth of using DL to significantly improve the state-of-the-art in many applications, particularly image, text classification, and speech recognition. 

\textbf{The Need for Hardware Acceleration:} Vast amounts of data powered by the exponential increase in computing capabilities have been instrumental in the success of DL. More notably, with the advent of the powerful Graphic Processing Unit (GPU) \citep{owens2008gpu}, training processes of the DL models have been drastically accelerated.  

Fast Matrix Multiplication has been heavily researched for the past several decades. We are now reaching a limit on speeding up matrix multiplication further. Furthermore, the need for astronomical size neural networks and unprecedented growth in the data volumes have worsened this problem. As a result, the community is heavily investing in dedicated hardware to take DL further beyond this point~\citep{jouppi2017datacenter}. Designing dedicated hardware is risky because they require significant investment and time to develop. Moreover, dedicated hardware caters to a specific algorithm for which they are designed. Thus, change in the state-of-the-art algorithms can render specialized hardware less effective in the future. However, for the case of DL, this investment is justified due to the lack of significant progress in the algorithmic alternatives for years. 

\textbf{Unsuccessful Alternatives to Matrix Multiplication:} On the orthogonal side, there have been several works on replacing the costly matrix multiplication with cheaper algorithmic alternatives ~\citep{le2014powers}. Unfortunately, we have seen minimal practical benefits from the algorithmic front. So far, there has been no demonstration, even remotely, that a smart algorithmic implementation in any form can outperform the advantages of hardware acceleration.  

\textbf{Exploiting Adaptive Sparsity in Neural Networks:} In popular frameworks like Tensorflow (TF), Sampled Softmax \citep{jean2015using} is deployed to approximate the full softmax efficiently. While sampled softmax offers computational savings, it has high estimation bias \citep{blanc2018adaptive}. This leads to poor convergence behavior which is empirically verified in our experiments in section \ref{sec:experiments}. In this paper, we will exploit the idea of adaptive sparsity \citep{blanc2018adaptive} or adaptive dropouts~\citep{ba2013adaptive}. The idea stems from several recent observations~\citep{makhzani2015winner, makhzani2013k} that we can accurately train neural networks by selectively sparsifying most of the neurons, based on their activation, during every gradient update. ~\citep{srivastava2014dropout} has also shown that selective sparsification can in-fact be superior in accuracy due to implicit regularization. However, selective sparsification does not directly lead to computational savings. ~\citep{spring2017scalable} shows the first possibility of an algorithmically efficient solution by employing Locality Sensitive Hash (LSH) tables to identify a sparse set of neurons efficiently during each update. The proposed algorithm has an added advantage of making the gradient update HOGWILD style parallel~\citep{recht2011hogwild}. Such parallelism does not hurt convergence because extremely sparse and independent updates are unlikely to overlap and cause conflicts of considerable magnitude. Despite all the niceness presented, current implementation of~\citep{spring2017scalable} fails to demonstrate that the computational advantage can be translated into a faster implementation when directly compared with hardware acceleration of matrix multiplication. In particular, it is not clear if we can design a system that can effectively leverage the computational advantage and at the same time compensate for the hash table overheads using limited (only a few cores) parallelisms. In this paper, we provide the first such implementation for large fully connected neural networks.

\subsection{Our Contributions}
Our main contributions are as follows:
\begin{itemize}[leftmargin=*,nosep,nolistsep]
\item We show the first C++ OpenMP based system SLIDE with modest multi-core parallelism on a standard CPU that can outperform the massive parallelism of a powerful V100 GPU on a head-to-head time-vs-accuracy comparison. 
This unique possibility is because the parallelism in SLIDE is naturally asynchronous by design. We have our code and benchmark scripts for reproducibility\footnote{https://github.com/keroro824/HashingDeepLearning}.
\item  We make several novel algorithmic and data-structural choices in designing the LSH based sparsification to minimize the computational overheads to a few memory lookups only (truly $O(1)$). At the same time, it does not affect the convergence of the DL algorithm. The implementation further takes advantage of the sparse gradient updates to achieve negligible update conflicts, which creates ideal settings for Asynchronous SGD (Stochastic Gradient Descent)~\citep{recht2011hogwild}. These contributions could be of independent interest in both the LSH and DL literature.
\item  We provide a rigorous evaluation of our system on two large benchmark datasets involving fully connected networks. We show that SLIDE, on a modest CPU can be up to 2.7x faster, in wall clock time, than the best possible alternative with the best possible choice of hardware, at any accuracy. We perform a CPU-efficiency analysis of SLIDE using Intel VTune Performance Analyzer and show that memory-bound inefficiencies reduce for SLIDE with an increasing number of cores while it is the opposite for TF-CPU.
\item  Our analysis suggests that SLIDE is a memory-bound application, prone to some bottlenecks described in appendix~\ref{sec:cache_thrashing}. With careful workload and cache optimizations (eg. Transparent Hugepages) and a data access pattern (eg. SIMD instructions), we further speed up SLIDE by roughly 1.3x, making the overall speed up to 3.5x faster than TF-GPU and over 10x faster than TF-CPU.
\end{itemize}

\section{Locality Sensitive Hashing}\label{sec:background}
Our paper is based on several recent and classical ideas in Locality Sensitive Hashing (LSH) and adaptive dropouts in neural networks. LSH is a family of functions with the property that similar input objects in the domain of these functions have a higher probability of colliding in the range space than non-similar ones. A popular technique for approximate nearest-neighbor search uses the underlying theory of \emph{Locality Sensitive Hashing}~\citep{indyk1998approximate}. In formal terms, consider $\mathcal{H}$ to be a family of hash functions mapping $\mathbb{R}^{D}$ to some set $\mathcal{S}$.
\begin{defn} [\bf LSH Family]\ A family $\mathcal{H}$ is called\\
    $(S_0,cS_0,p_1,p_2)$-sensitive if for any two points $x,y \in \mathbb{R}^{D}$ and $h$ chosen uniformly from $\mathcal{H}$ satisfies the following:
    \begin{itemize}[leftmargin=*,nosep,nolistsep]
        \item if $Sim(x,y)\ge S_0$ then ${Pr}(h(x) = h(y)) \ge p_1$
        \item if $ Sim(x,y)\le cS_0$ then ${Pr}(h(x) = h(y)) \le p_2$
    \end{itemize}
    \label{def:lsh}
\end{defn}
Typically, for approximate nearest-neighbor search, we need $p_1 > p_2$ and $c < 1$ to hold. An LSH allows us to construct data structures that give provably efficient query time algorithms for the approximate nearest-neighbor problem with the associated similarity measure.

One sufficient condition for a hash family $\mathcal{H}$ to be an LSH family is that the \emph{\bf collision probability} ${Pr}_\mathcal{H}(h(x) = h(y))$ should be a monotonically increasing with the similarity, i.e. \begin{equation}\label{eq:monotonic}{Pr}_\mathcal{H}(h(x) = h(y)) = f(Sim(x,y)),\end{equation} where f is a monotonically increasing function. In fact, most of the popular known LSH families, such as Simhash \citep{gionis1999similarity} and WTA hash \citep{yagnik2011power, chen2018densified}, satisfy this strong property. It can be noted that Equation~\ref{eq:monotonic} automatically guarantees the two required conditions in the Definition~\ref{def:lsh}.

It was shown in~\citep{indyk1998approximate} that having an LSH family for a given similarity measure is sufficient for efficiently solving nearest-neighbor search in sub-linear time.
% \begin{thm} Given a family of
%     $(S_0, cS_0, p_1, p_2)$-sensitive hash functions, one can construct a data
%     structure for c-NN with $O(n^\rho \log n)$ query time and space $O(n^{1+\rho})$, where $\rho = \frac{\log p_1}{ \log p_2}< 1$.
%     \label{def:lshquery}
% \end{thm}

\textbf{The Algorithm:} The LSH algorithm uses two parameters, $(K, L)$. We construct $L$ independent hash tables. Each hash table has a meta-hash function $H$ that is formed by concatenating $K$ random independent hash functions from the collection $\mathcal{F}$. Given a query, we collect one bucket from each hash table and return the union of $L$ buckets. Intuitively, the meta-hash function makes the buckets sparse and reduces the number of false positives, because only valid nearest-neighbor items are likely to match all $K$ hash values for a given query. The union of the $L$ buckets decreases the number of false negatives by increasing the number of potential buckets that could hold valid nearest-neighbor items. The candidate generation algorithm works in two phases [See \citep{spring2017new} for details]:
\begin{enumerate}[leftmargin=*,nosep,nolistsep]
    \item {\bf Pre-processing Phase:} We construct $L$ hash tables from the data by storing all elements $x$. We only store pointers to the vector in the hash tables because storing whole data vectors is very memory inefficient.
    \item {\bf Query Phase:} Given a query $Q$; we search for its nearest-neighbors. We report the union from all of the buckets collected from the $L$ hash tables. Note that we do not scan all the elements but only probe $L$ different buckets, one bucket for each hash table.
\end{enumerate}
After generating the set of potential candidates, the nearest-neighbor is computed by comparing the distance between each item in the candidate set and the query.
\begin{figure}[t]
 \vspace{-1mm}
    \centering
    \mbox{
        \includegraphics[width=2.0in]{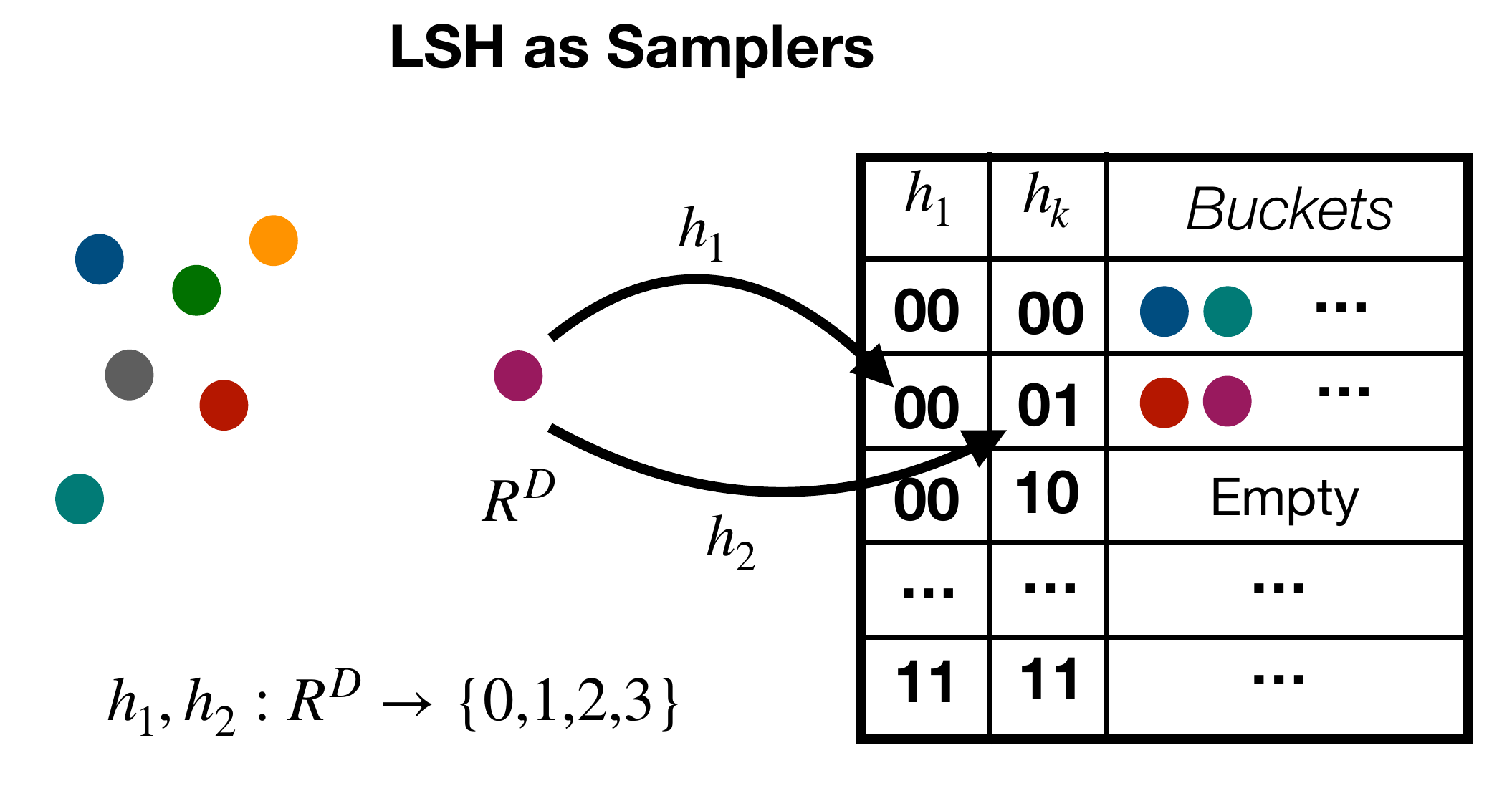}
    }
    \vspace{-2mm}
    \caption{Schematic diagram of LSH. For an input, we obtain hash codes and retrieve candidates from the respective buckets.}
    \label{fig:lsh}
    \vspace{-3mm}
\end{figure}
\subsection{LSH for Estimation and Sampling}\label{lshe}
Although LSH provides provably fast retrieval in sub-linear time, it is known to be very slow for accurate search because it requires very large number of tables, i.e. large $L$. Also, reducing the overhead of bucket aggregation and candidate filtering is a problem on its own. Consequent research led to the sampling view of LSH~\citep{spring2017scalable,chen2018unique,chen2019fast,luo2018scaling} that alleviates costly searching by efficient sampling, as shown in figure~\ref{fig:lsh}. It turns out that merely probing a few hash buckets (as low as 1) is sufficient for adaptive sampling. Observe that an item returned as a candidate from a $(K,L)$-parameterized LSH algorithm is sampled with probability $1 - (1 - p^K)^L$, where $p$ is the collision probability of LSH function (sampling probability is monotonic in $p$). Thus, with LSH algorithm, the candidate set is adaptively sampled where the sampling probability changes with $K$ and $L$.

This sampling view of LSH was the key for the algorithm proposed in paper~\citep{spring2017scalable} that shows the first possibility of adaptive dropouts in near-constant time, leading to efficient backpropagation algorithm.    

%A year later, ~\citep{spring2017new} demonstrated the first theory of using these samples for unbiased estimation of partition functions in log-linear models. More specifically, the authors showed that since we know the precise probability of sampled elements  $1 - (1 - p^K)^L$, we could design provably unbiased estimators using importance sampling type idea. This was the first demonstration that random sampling could be beaten with roughly the same computational cost as vanilla sampling. ~\citep{luo2017arrays} used the same approach for unbiased estimation of anomaly scoring function. ~\citep{charikarhashing} rigorously formalized these notions and showed provable improvements in sample complexity of kernel density estimation problems. Recently, ~\citep{chen2017unique} used the sampling in a very different context of connected component estimation for unique entity counts.

\subsubsection{MIPS Sampling}
Recent advances in maximum inner product search (MIPS) using asymmetric locality sensitive hashing has made it possible to sample large inner products. Given a collection $\mathcal{C}$ of vectors and query vector $Q$, using $(K,L)$-parameterized LSH algorithm with MIPS hashing~\citep{shrivastava2014asymmetric}, we get a candidate set $S$. Every element in $x_i \in \mathcal{C}$ gets sampled into $S$ with probability $p_i$, where $p_i$ is a monotonically increasing function of $Q \cdot x_i$. Thus, we can pay a one-time linear cost of preprocessing $\mathcal{C}$ into hash tables, and any further adaptive sampling for query $Q$ only requires few hash lookups.

\begin{figure*}[t!]
    \vspace{-1mm}
    \centering
    \includegraphics[width=6in]{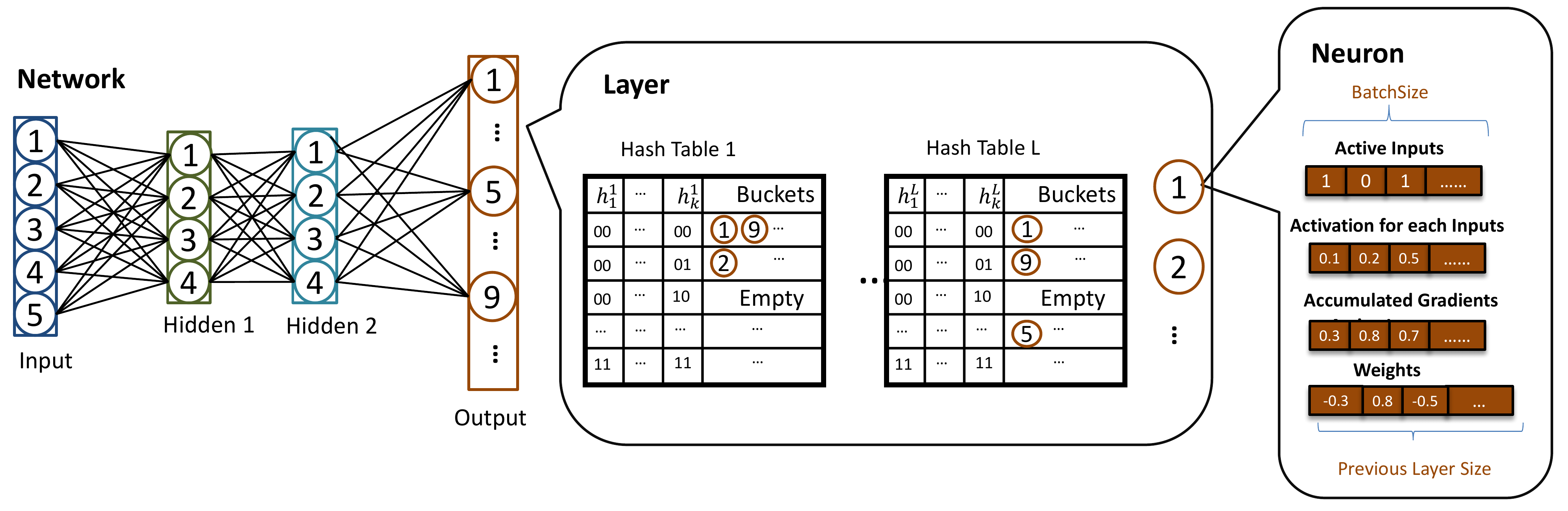}
     \vspace{-3mm}
    \caption{Architecture: The central module of SLIDE is Network. The network is composed of few-layer modules. Each layer module is composed of neurons and a few hash tables into which the neuron ids are hashed. Each neuron module has multiple arrays of batch size length: 1) a binary array suggesting whether this neuron is active for each input in the batch 2) activation for each input in the batch 3) accumulated gradients for each input in the batch. 4) The connection weights to the previous layer. The last array has a length equal to the number of neurons in the previous layer.}
    \label{fig:workflow}
     \vspace{-3mm}
\end{figure*}
% \begin{minipage}{0.5\textwidth}

% \end{minipage}

\begin{algorithm}[h!]
\begin{algorithmic}[1]
    \STATE \textbf{Input: data $X$, iterations $n$, batch size $B$}
    \STATE \textbf{Output: $\theta$}
    \STATE Initialize weights $w_l$ for each layer $l$
    \STATE Create hash tables $HT_l$, functions $h_l$ for each layer $l$
    \STATE Compute $h_l(w_l^a)$ for all neurons
    \STATE Insert all neuron ids $a$, into $HT_l$ according to  $ h_l(w_l^a)$
    \FOR {$i=1:n$}
    \STATE $\text{Input}_0$ = Batch$(X, B)$
    \FOR {$l=1:$ Layers}
    \STATE $S_l$ = Sample$(\text{Input}_{l-1}, HT_l, h_l)$  (Algorithm~\ref{algo2}) 
    \STATE Activation = Forward Propagation ($\text{Input}_{l-1}$, $S_l$)
    \STATE $\text{Input}_{l} =$ Activation
    \ENDFOR
    \FOR {$l=1:$ Layers}
    \STATE Backpropagation ($S_l$)
    \ENDFOR
    \ENDFOR
    \STATE return $\theta$
\end{algorithmic}
\caption{SLIDE Algorithm}\label{algo1}
\end{algorithm}
\begin{algorithm}[h!]
    \begin{algorithmic}[1]
        \STATE \textbf{Input: $Q$, $HT$, $h$ }
        \STATE \textbf{Output: $S_l$ (a set of active neurons on layer $l$)}
        \STATE Compute $h(Q)$.
        \FOR {$t=1:L$}
        \STATE $S$ = $S \cap$ Query($h_l(Q_l)$, $HT_l^t$)
        \ENDFOR
        \STATE return $S$
    \end{algorithmic}
    \caption{Algorithm for LSH Sampling}\label{algo2}
\end{algorithm}

\section{Proposed System: SLIDE}

\subsection{Introduction to the overall system}\label{subsec:design}

Before introducing SLIDE in details, we define important notations:
{\bf 1)} $B$: input batch size
{\bf 2)} $N_l^j$: Neuron $j$ in layer $l$
{\bf 3)} $x_l$: inputs for layer $l$ in the network
{\bf 4)} $w_l^a$: weights for $a^{th}$ neuron in layer $l$
{\bf 5)} $h_l$: hash functions in layer $l$
{\bf 6)} $\bold{N_l^a}$:  the set of active neurons in layer $l$ for the current input.

\textbf{Initialization:} Figure \ref{fig:workflow} shows the modular structure of SLIDE and algorithm~\ref{algo1} shows the detailed steps. Every layer object contains a list of neurons and a set of LSH sampling hash tables. Each hash table contains ids of the neurons that are hashed into the buckets.  During the network initialization, the weights of the network are initialized randomly. Afterwards, $K\times L$ LSH hash functions are initialized along with $L$ hash tables for each of the layers. For instance, the example network in Figure \ref{fig:workflow} maintains hash tables in two hidden layers as well as the output layer. The details of using various hash functions are discussed in appendix~\ref{sec:hashfunctions}. The LSH hash codes $h_l(w_l^a)$ of the weight vectors of neurons in the given layer are computed according to the hash functions. The id $a$ of the neuron is saved into the hash buckets mapped by the LSH function $h_l(w_l^a)$. This construction of LSH hash tables in each layer is a one-time operation which can easily be parallelized with multiple threads over different neurons in the layer independently.  
\begin{figure}[h!]
\vspace{-1mm}
    \centering
    \includegraphics[width=2in]{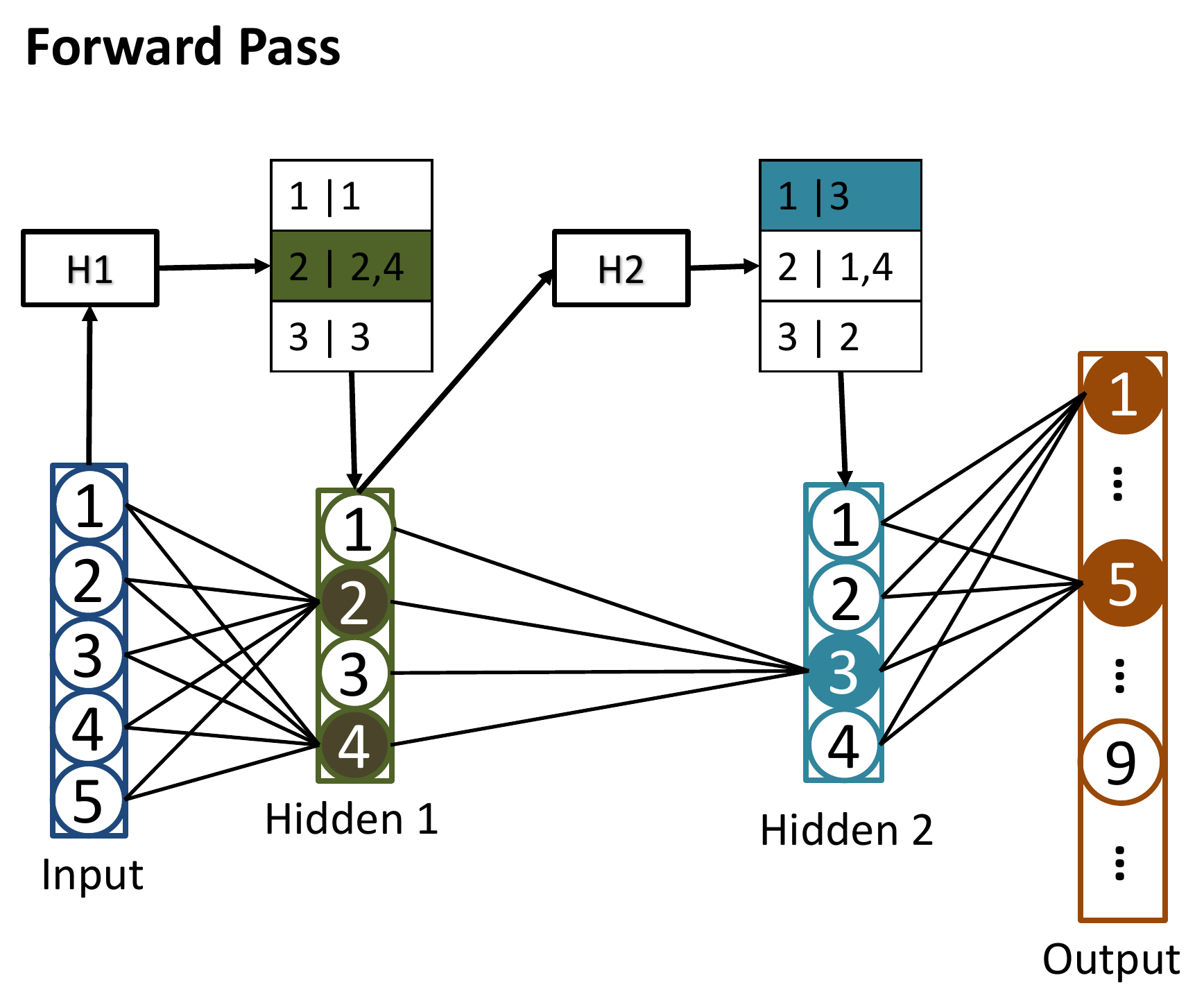}
    \vspace{-3mm}
    \caption{Forward Pass: Given an input, we first get the hash code $H1$ for the input, query the hash table for the first hidden layer, and obtain the active neurons. We get the activations for only this set of active neurons. We do the same for the subsequent layers and obtain a final sparse output. In practice, we use multiple hash tables per layer.}
        \label{fig:forward}
        \vspace{-5mm}
\end{figure}

\textbf{Sparse Feed-Forward Pass with Hash Table Sampling:} In the feed-forward phase, given a single training instance, we compute the network activation until the final layer, which gives us the output. In SLIDE, instead of calculating all the activations in each layer, the input to each layer $x_l$ is fed into hash functions to compute $h_l(x_l)$. The hash codes serve as a query to retrieve ids of active (or sampled) neurons from the matching buckets in hash tables. For example, in the figure~\ref{fig:forward}, $h_1(x_1)$ is first computed and then used to retrieve $N_1^2$ and $N_1^4$ as the active neurons. Only the activations of active neurons are calculated and passed on as the inputs to the next layer. The other activations, like those of $N_1^1$ and $N_1^3$, are directly treated as $0$ and never computed. We describe our design choices that reduce the sampling overheads significantly in section \ref{sec:sampling}.

The above-described operations are performed sequentially in every layer, starting from the very first layer where the input is the data itself. Even in the output layer, which has softmax activation, only neurons sampled from hash tables are treated as active neurons. For softmax, for every active neuron, we compute its output as $\sigma(N_{o}^k)  = \frac{e^{x_{o}w_{o}^k}}{\sum_{\bold{N_{o}^a}}e^{x_{o}w_{o}^k}}.$ Note that the normalizing constant for softmax is no longer the sum over all neurons but only the active ones. 

\textbf{Sparse Backpropagation or Gradient Update:}
The backpropagation step follows the feed-forward step. After computing the output of the network, we compare it with the known label of the input and backpropagate the errors layer-by-layer to calculate the gradient and update the weights. Here we use the classical backpropagation message passing type implementation rather than vector multiplication based. For every training data instance, after updating the weights of any given neuron, the neuron propagates the partial gradients (using error propagation) back to only active neurons in previous layers via the connected weights. As a result, we never access any non-active neuron or any non-active weight, which is not part of the feed-forward process on a given input. This process ensures that we take full advantage of sparsity. Our computation over each input is only of the order of active neurons and weights rather than the total number of parameters. It should be noted that if we compute activation for $s < 1$ fraction of neurons in each layer (on an average), the fraction of weights that needs to  be updated is $s^2$ only, which is a significant reduction when $s$ is small (as is the case for our experiments).

\vspace{-1mm}
\textbf{Update Hash Tables after Weight Updates:} After the weights are updated, we need to modify the positions of neurons in the hash tables accordingly. Updating neurons typically involves deletion from the old bucket followed by an addition to the new bucket, which can be expensive. We discuss several design tricks that we use to overcome this overhead of updating hash tables in section~\ref{subsec:hashtable}.

\vspace{-1mm}
\textbf{OpenMP Parallelization across a Batch:} 
For any given training instance, both the feed-forward and backpropagation operations are sequential as they need to be performed layer by layer. 
% The clear advantage of SLIDE is that the total arithmetic operation due to extreme sparsity of neurons is notably less than the matrix multiplication operation. All operations are performed in a sparse fashion, where weights, layers, and neurons are accessed by their ids.  Values of zeros are never involved in any memory accesses or computations. 
SLIDE uses usual Batch Gradient Descent with Adam optimizer, where the batch size is generally in the order of hundreds.  Each data instance in the batch runs in a separate thread and its gradients are computed in parallel. To ensure the independence of computation across different threads, every neuron stores two additional arrays, each of whose length is equal to the batch size. These arrays keep track of the input specific neuron activations and error gradients. Every input is assigned an id, which can be used as an index to locate its activation (or error gradient) on any neuron. Besides, we also have a bit array at each neuron to determine whether the particular input activates a neuron or not. This small memory overhead is negligible for CPUs but it ensures that the gradient computation is independent across different instances in the batch.

\vspace{-0.5mm}
The extreme sparsity and randomness in gradient updates allow us to asynchronously parallelize the accumulation step of the gradient across different training data without leading to a considerable amount of overlapping updates. SLIDE heavily capitalizes on the theory of HOGWILD~\citep{recht2011hogwild}, which shows that a small amount of overlap is tolerable. It does not hurt the convergence even if we resolve the concurrent updates randomly. Thus, after independently computing the gradients, each thread pushes the updates directly to the weights asynchronously. This asynchronous update avoids synchronization during batch accumulation which is otherwise sequential in the batch. 

\vspace{-0.5mm}
In section~\ref{sec:scalability_tests}, we observe that due to this asynchronous choice, we obtain near-perfect scaling of our implementation with an increasing number of cores. Such perfect scaling is particularly exciting because even highly optimized implementation of TF on CPUs shows poor scaling behavior with increasing cores beyond 16 cores.    

\subsection{Details of Hash Functions and Hash Tables}\label{sec:function}

SLIDE provides a natural trade-off between the efficiency of retrieving active neurons and the quality of the retrieved ones. To facilitate this, we have three tunable parameters $K, L, B$. As mentioned in section~\ref{sec:background}, $L$ serves as the number of hash tables. To determine which bucket to choose, we use $K$ hash codes for each hash table. Hence, SLIDE generates $K \times L$ randomized hash functions all belonging to one hash family for each layer. In every bucket in a hash table, the number of entries is limited to a fixed bucket size. Such a limit helps with the memory usage and also balances the load on threads during parallel aggregation of neurons.

% As mentioned in Section \ref{sec:background}, hash codes that are used to determine which bucket to choose in each hash table are produced by $K$ randomized hash functions which all belong to one hash family. Disregarding the type of hash functions, the number of buckets or the length of hash tables is fixed initially and serves as a hyper-parameter of the system. Furthermore, the bucket size $B$ and the number of hash tables $L$ are also adjustable. Therefore, $K$, $L$ and $B$ are the three tunable and customized parameters in every layer of the network to control the quality and efficiency of the active neuron retrieval.

In our implementation of SLIDE, we support four types of hash functions from LSH family: 1) Simhash 2) WTA hash 3) DWTA hash and 4) Minhash respectively. Each of these hash families preserves different similarities and hence is useful for various scenarios. We discuss the implementation details of Simhash and DWTA hash below and others in appendix \ref{sec:hashfunctions}. SLIDE also provides the interface to add customized hash functions based on need.
% \vspace{-0.5cm}

\textbf{Simhash \citep{gionis1999similarity}: }SimHash is a popular LSH for the {\it cosine} similarity measure. We use $K \times L$ number of random pre-generated vectors with components taking only three values  $\{+1, 0, -1\}$. The reason behind using only $+1s$ and $-1s$ is for fast implementation. It requires additions rather than multiplications, thereby reducing the computation and speeding up the hashing process. 
To further optimize the cost of Simhash in practice, we can adopt the sparse random projection idea \citep{li2006very}. %The basic idea is to shrink the dimension of the random vectors. However, this creates a mismatch between the dimension of the input and the random vectors while computing the inner product. 
A simple implementation is to treat the random vectors as sparse vectors and store their nonzero indices in addition to the signs. For instance,  let the input vector for Simhash be in $R^d$. Suppose we want to maintain $1/3$ sparsity, we may uniformly generate $K*L$ set of $d/3$ indices from $[0, d-1]$. In this way, the number of multiplications for one inner product operation during the generation of the hash codes would simply reduce from $d$ to $d/3$. Since the random indices are produced from one-time generation, the cost can be ignored. 

\textbf{DWTA hash \citep{chen2018densified}: } DWTA hash transforms the input feature space into binary codes such that the Hamming distance in the resulting space closely correlates with rank similarity measure for sparse data. We generate $\frac{KLm}{d}$ number of permutations and every permutation is split into $\frac{d}{m}$ bins. DWTA loops through all the nonzero (NNZ) indices of the sparse input. For each of them, we update the current maximum index of the corresponding bins according to the mapping in each permutation. It should be noted that the number of comparisons and memory lookups in this step is $O(NNZ*\frac{KLm}{d})$, which is significantly more efficient than simply applying WTA hash to sparse input. For empty bins, the densification scheme proposed in \citep{chen2018densified} is applied.

\section{Reducing Overhead}
\subsection{Sampling Overhead}\label{sec:sampling}
The key idea of using LSH for adaptive sampling of neurons is sketched in section \ref{subsec:design}. We have designed three strategies to sample neurons with large activation: 1) Vanilla Sampling 2) Topk Sampling 3) Hard Thresholding. We introduce them here and discuss their utility and efficiency in appendix~\ref{sec:sampling2}. 
% Experiments comparing these strategies are reported in section \ref{sec:design_choices} in supplementary.

{\bf Vanilla Sampling: } Denote $\beta_l$ as the number of active neurons we target to retrieve in layer $l$. After computing the hash codes of the input, we randomly choose a table and only retrieve the neurons in its corresponding bucket. We continue retrieving neurons from another random table until $\beta_l$ neurons are selected or all the tables have been looked up. Let us assume we retrieve from $\tau$ tables in total. Formally, the probability that a neuron $N_l^j$ gets chosen is, $Pr(N_l^j) = (p^K)^{\tau}(1-p^K)^{L-\tau},$
where $p$ is the collision probability of the LSH function that SLIDE uses. The time complexity of vanilla sampling is $O(\beta_l)$.

{\bf TopK Sampling: } In this strategy, the basic idea is to obtain those neurons that occur more frequently among all $L$ hash tables. After querying with the input, we first retrieve all the neurons from the corresponding bucket in each hash table and aggregate their frequencies across all hash tables. The frequencies are sorted, and only the neurons with top $\beta_l$ frequencies are selected. 
This requires additional $O(\vert N_l^a \vert)$ space for maintaining the hashmap and $O(\vert N_l^a \vert+ \vert N_l^a \vert log \vert N_l^a \vert)$ time for both sampling and sorting. 

{\bf Hard Thresholding:} In this strategy, we bypass the sorting step in TopK sampling by selecting neurons that appear at least $m$ times in the retrieved buckets. Here, the probability that a neuron $N_l^j$ gets chosen is,
$Pr(N_l^j) = \sum_{i = m}^{L} \tbinom {L}{i}(p^K)^{i}(1-p^K)^{L-i}$.
% Figure \ref{fig:thres} shows a sweep of curves that present the relation between collision probability of $h_l(w_l^j)$ and $h_l(x_l)$  and the probability that neuron $N_l^j$ is selected under various values of $m$ when $L=10$. 
% we can visualize the trade-off between collecting more good neurons and omitting bad neurons by tweaking $m$ in figure \ref{fig:thres}. For a high threshold like $m=9$, only the neurons with $p>0.8$ have more than $Pr>0.5$ chance of retrieval. This ensures that bad neurons are eliminated but the retrieved set might be insufficient. However, for a low threshold like $m=1$, all good neurons are collected but bad neurons with $p<0.2$ are also collected with $Pr>0.8$. Therefore, depending on the tolerance for bad neurons, we choose an intermediate $m$ in practice.  

Figure \ref{fig:sort} is a preview of the empirical efficiency comparison of above three strategies shown in appendix~\ref{sec:sampling2}. We see that Vanilla sampling is a lot more time efficient than the other two strategies at the cost of sample quality.

% \begin{figure}[tb]
% 	\centering
% 	\mbox{
% 		\includegraphics[width=2in]{pic/spec.eps}
% 	}
% 	\caption{Hard Thresholding: Theoretical selection probability $Pr$ vs the collision probabilities $p$ for various values of frequency threshold $m$. High threshold ($m=9$) gets less number of false positive neurons but misses out on many active neurons. A low threshold ($m=1$) would select most of the active neurons along with lot of false positives.}
% 	\label{fig:thres}
% \end{figure}

\begin{figure}[t]
% 	\label{softmax}
	   % \vspace{-2mm}
	\centering
	\mbox{
		\includegraphics[width=2.3in]{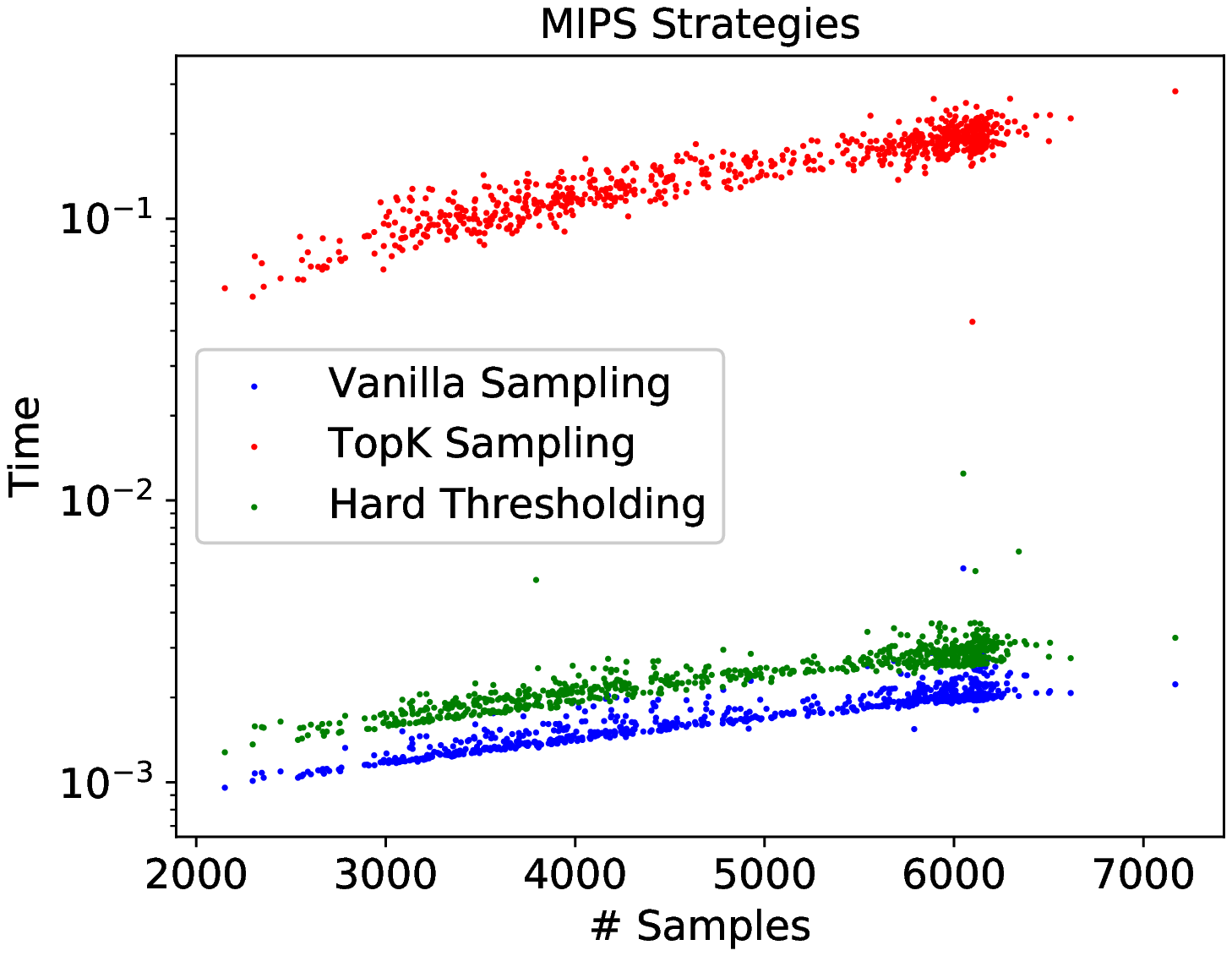}
	}
	 \vspace{-2mm}
	\caption{Time consumed (in seconds) for various sampling strategies for retrieving active neurons from hash tables.}
	\label{fig:sort}
	\vspace{-1cm}
\end{figure}

\subsection{Updating Overhead}\label{subsec:hashtable}
We introduce the following heuristics for addressing the expensive costs of updating the hash tables:

{\bf 1)} Recomputing the hash codes after every gradient update is computationally very expensive. Therefore, we dynamically change the update frequency of hash tables to reduce the overhead. Assume that we update the hash tables for the first time after $N_0$ iterations. Let $t-1$ be the number of times the hash tables have already been updated. We apply exponential decay on the update frequency such that the $t^{th}$ hash table update happens on iteration $\sum_{i=0}^{t-1} N_0e^{\lambda i}$, where $\lambda$ is a tunable decay constant. The intuition behind this scheme is that the gradient updates in the initial stage of the training are larger than those in the later stage, especially while close to convergence.

{\bf 2)} SLIDE needs a policy for adding a new neuron to a bucket when it is already full. To solve such a problem, we use the same solution in \citep{wang2018randomized} that makes use of Vitter’s reservoir sampling algorithm \citep{vitter1985random} as the replacement strategy. It was shown that reservoir sampling retains the adaptive sampling property of LSH tables, making the process sound. Additionally, we implement a simpler alternative policy based on FIFO (First In First Out).

{\bf 3)}  For Simhash, the hash codes are computed by $h^{sign}_w(x) = sign(w^Tx)$. During backpropagation, only the weights connecting the active neurons across layers get updated. Only those weights contribute to the change of $w^Tx$. Therefore, we can also memorize the result of $w^Tx$ besides the hash codes.  When $x \in R^d$ gets updated in only $d^{'}$ out of $d$ dimensions, where $d^{'} \ll d$, we only need $O(d^{'})$ rather than $O(d)$ addition operations to compute the new hash codes.

\section{Evaluations}\label{sec:experiments}

In this section, we're going to empirically investigate SLIDE's performance on multiple fronts such as: {\bf 1)} SLIDE against TF-GPU with V100s {\bf 2)} SLIDE against TF-CPU {\bf 3)} SLIDE's adaptive sampling against sampled softmax (plain random sampling) {\bf 4)} Scalability against TF-CPU with CPU core count {\bf 5)} Effect of batch size {\bf 6)} Benefits of Design Choices. While we focus on evaluating the basic aspects of SLIDE, we additionally perform several CPU optimizations like support for Kernel Hugepages to reduce cache misses which improve SLIDE's performance by $\approx 30\%$. The optimization details are given in appendix~\ref{sec:HPC} and the improvement in performance is shown in section \ref{sec:Hugepages}.

\begin{table}
    \label{table:data}
    \centering
    \vspace{-4mm}
    \caption{Statistics of the datasets}
    \resizebox{\linewidth}{!}{
    \begin{tabular}{|c|c|c|c|c|l|} \hline
    	& Feature Dim & Feature Sparsity & Label Dim & Training Size & Testing Size \\ \hline
    	Delicious-200K & 782,585 & 0.038 \% & 205,443 & 196,606 & 100,095 \\ \hline
    	Amazon-670K & 135,909 & 0.055 \% & 670,091 & 490,449 & 153,025\\
    	\hline\end{tabular}
    	}
    	\vspace{-4mm}
\end{table}

\begin{figure*}[tb]
    % \hspace{-.15in}
    \centering
    \mbox{
        \includegraphics[width=1.7in]{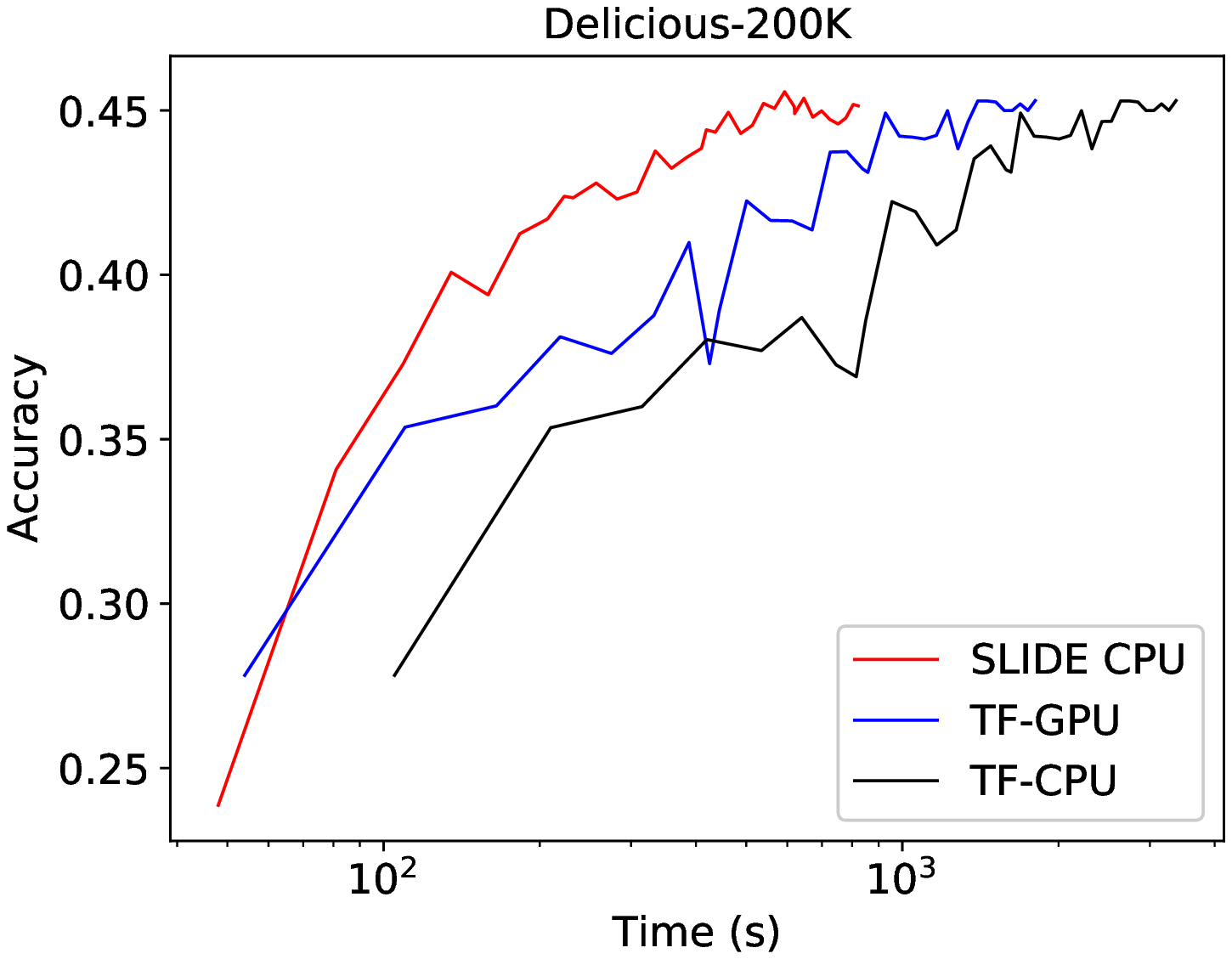}
        % \hspace{-.2in}
        \includegraphics[width=1.7in]{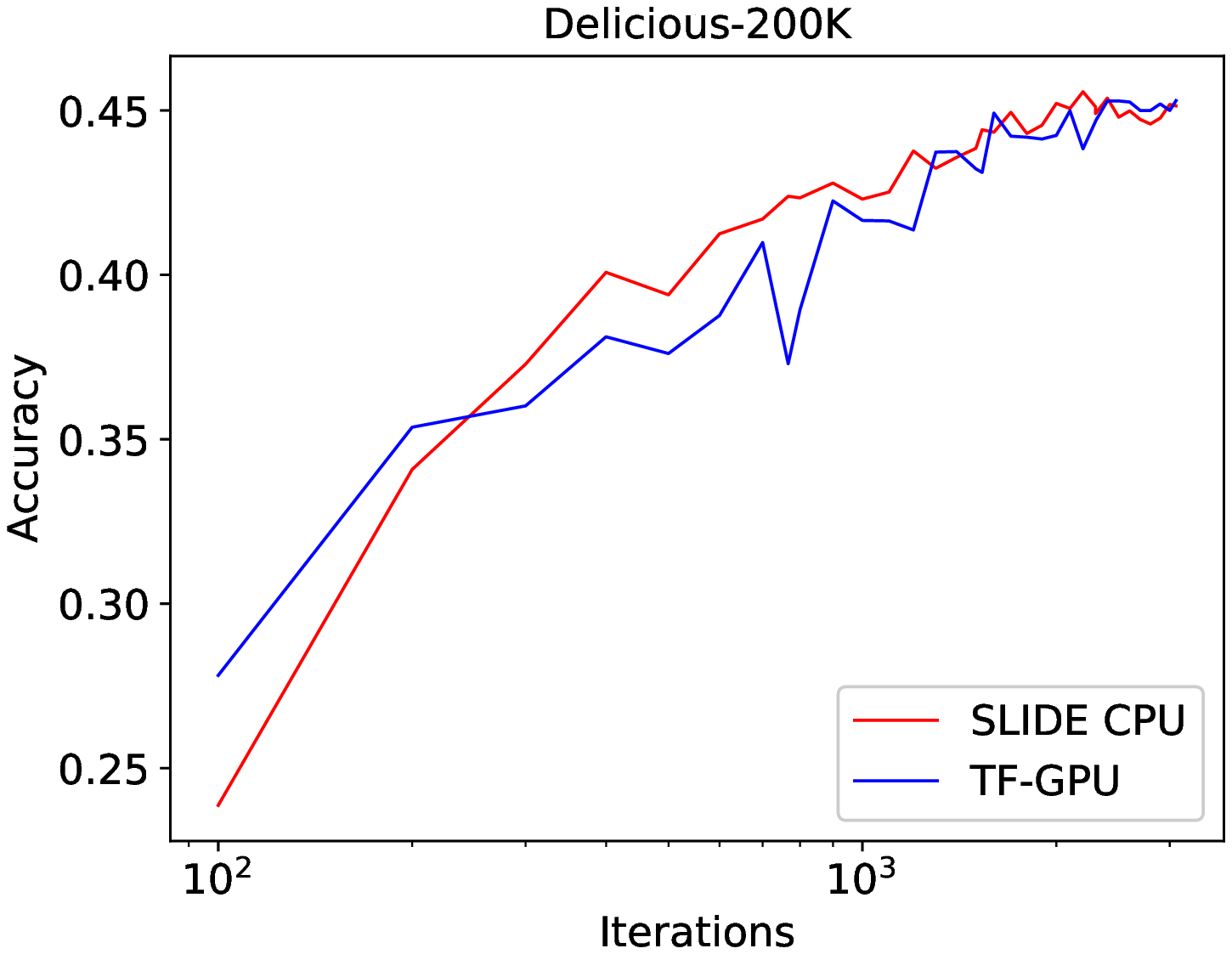}
        % \hspace{-.2in}
        \includegraphics[width=1.7in]{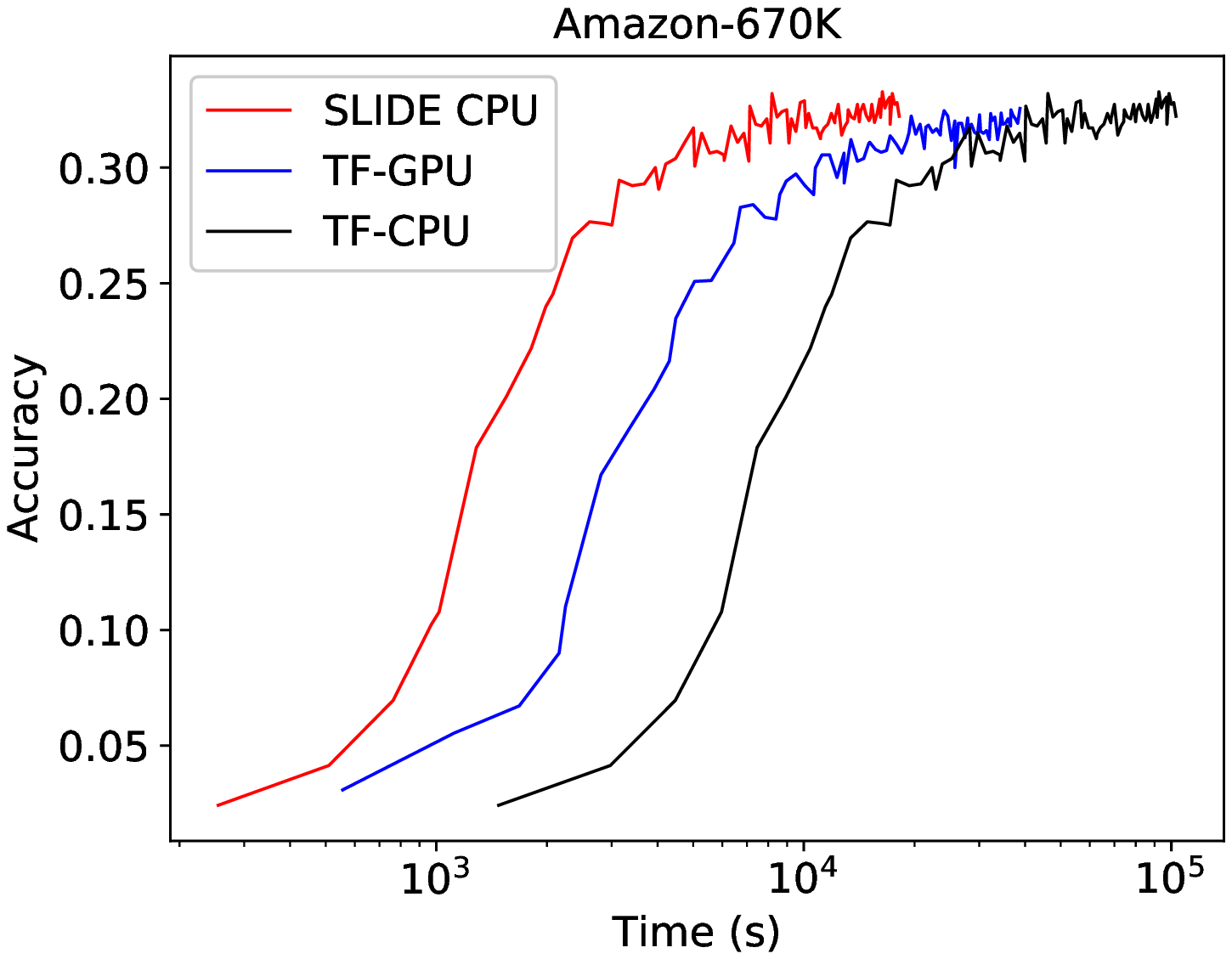}
        % \hspace{-.2in}
        \includegraphics[width=1.7in]{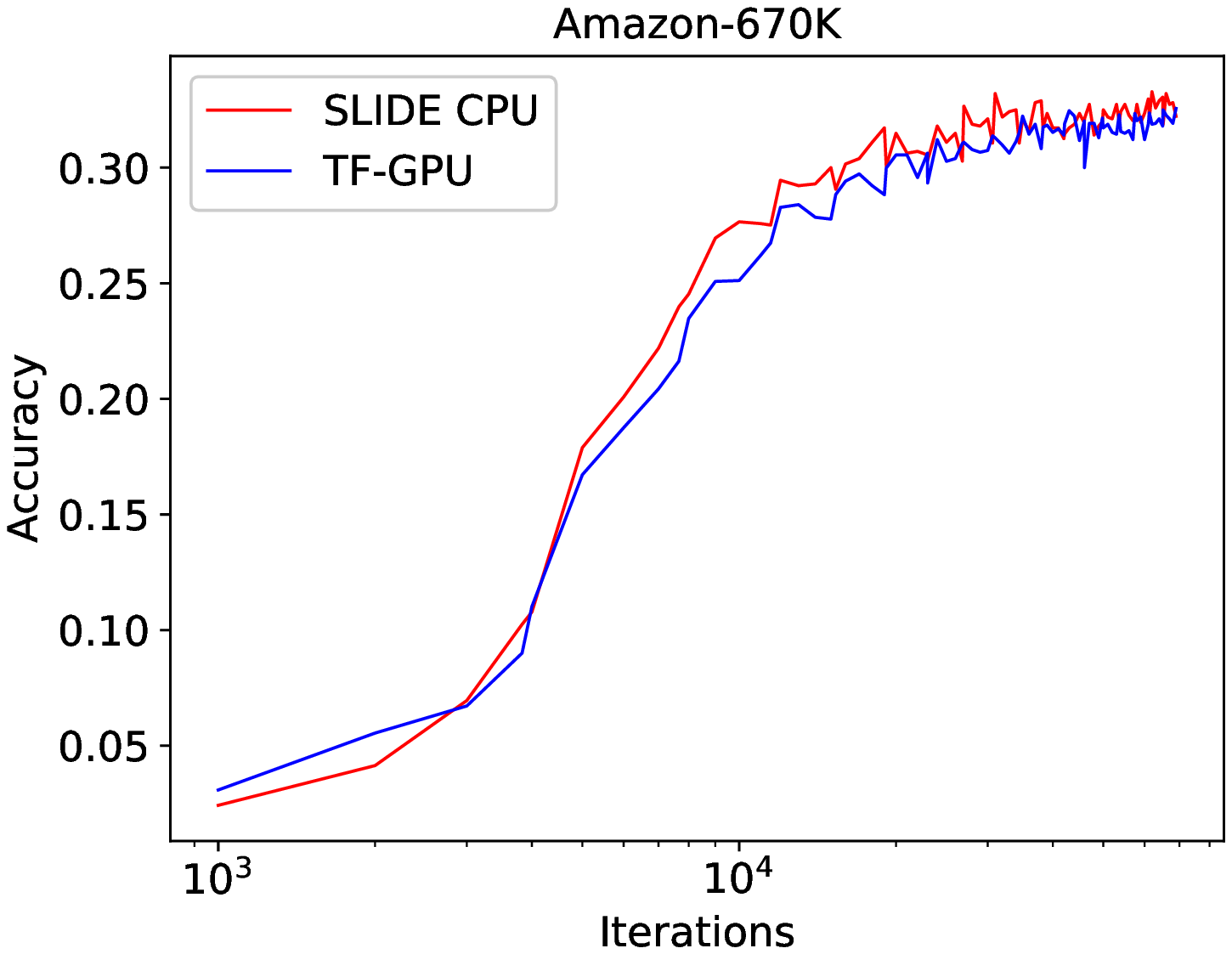}
        }
            \vspace{-3mm}
    \caption{It shows the comparison of SLIDE (in red) against TF-GPU (in blue) and TF-CPU (in black). The x-axis is plotted in log scale to accommodate the otherwise slow TF-CPU curve. We notice that the time required for convergence is 2.7x lower than that of TF-GPU. When compared against iterations, the convergence behavior is identical, which confirms that the superiority of SLIDE is due to algorithm and implementation and not due to any optimization bells and whistles.}
    \label{fig:full}
\end{figure*}

\textbf{Fully-Connected Large Architecture:} Fully connected networks are common in most applications. 
% and dominate most applications except vision where the use of convolutional neural networks (CNN) are more pronounced. Thus, our evaluation is limited to only fully connected architectures.
To show SLIDE's real advantage, we will need large networks where even a slight decrease in performance is noticeable. Thus, the publicly available extreme classification datasets, requiring more than 100 million parameters to train due to their extremely wide last layer, fit this setting appropriately. For these tasks, most of the computations (more than $99\%$) are in the final layer.

\textbf{Datasets:} We employ two large real datasets, Delicious-200K and Amazon-670K, from the Extreme Classification Repository \cite{extreme}. Delicious-200K dataset is a sub-sampled dataset generated from a vast corpus of almost 150 million bookmarks from Social Bookmarking Systems (del.icio.us). Amazon-670K dataset is a product to product recommendation dataset with 670K labels. The statistics of the datasets are included in Table~\ref{table:data}.

\textbf{Infrastructure:} All the experiments are conducted on a server equipped with two 22-core/44-thread processors (Intel Xeon E5-2699A v4 2.40GHz) and one NVIDIA Tesla V100 Volta 32GB GPU. The server has an Ubuntu 16.04.5 LTS system with the installation of TF-GPU 1.12. We compiled TF-CPU 1.12 from source with GCC5.4 in order to support FMA, AVX, AVX2, SSE4.1, and SSE4.2 instructions, which boost the performance of TF-CPU by about $35\%$. SLIDE is written in C++ and compiled under GCC5.4 with OpenMP flag. The most exciting part is that SLIDE only uses vanilla CPU thread parallelism and yet outperforms TF-GPU (V100) by a large margin in performance.

\textbf{Baselines:} We benchmark the tasks with our system SLIDE, and compare against highly optimized TF framework for both CPU and GPU. Specifically, the comparison is between the same tasks, with the exact same architecture, running on TF-CPU and TF-GPU.  The optimizer and the learning hyperparameters (details later) were also the same to avoid unfair comparisons.  
Most of the computations in our architecture are in the softmax layer. Hence, to corroborate the advantage of adaptive sampling~\citep{yen2018loss} vs vanilla sampling, we also compare against the popular sampled softmax algorithm~\citep{jean2015using} which is a fast proxy to the full softmax. We use the optimized Sampled Softmax functionality provided in TF-GPU. 
% In principle, both SLIDE and Sampled Softmax accelerate the training in the same way, i.e., by selecting a few neurons and passing gradients only from those neurons. While Sampled Softmax makes a naive static sampling of neurons, SLIDE uses adaptive sampling which is known to be superior in deep learning literature \citep{yen2018loss}. 
This comparison sheds light on the necessity of LSH based input dependent adaptive sampling compared to static sampling scheme which is the only other alternative in practice. 

\textbf{Hyper Parameters:} For both the datasets, we adopt the same model architecture in \citep{yen2018loss}. We choose the standard fully connected neural network with one hidden layer of size 128. We choose a batch size of 128 for Delicious-200K dataset and 256 for Amazon-670K dataset as the input dimension for the former is very large. We run all algorithms until convergence. To quantify the superiority of SLIDE over other baselines, we also use the same optimizer, Adam \citep{kingma2014adam} by varying the initial step size from $1e^{-5}$ to $1e^{-3}$ which leads to better convergence in all experiments. For SLIDE, we maintain the hash tables for the last layer, where we have a computational bottleneck of the models. For specific LSH setting, we choose Simhash, $K=9$, $L=50$ for Delicious dataset and DWTA hash, $K=8, L=50$ for Amazon-670k dataset. We update the hash tables with an initial update period of $N_0 = 50$ iterations and then exponentially decaying (section~\ref{subsec:hashtable}).

\textbf{Main Results:}\label{sec:results} We show the time and iteration wise comparisons for SLIDE vs TF GPU/CPU in Figure \ref{fig:full}. Note that the $x$-axis is in log-scale, and all the curves have a long flat converged portion when plotted on a linear scale indicating clear convergence behavior. Red, blue and black lines represent the performance of SLIDE, TF-GPU, TF-CPU, respectively. We can see from the plots that SLIDE on CPU achieves any accuracy faster than TF on V100. TF-GPU is always faster than TF-CPU which is expected. It should be noted that these datasets are very sparse, e.g., Delicious dataset has only 75 non-zeros on an average for input features, and hence the advantage of GPU over CPU is not always noticeable. %But V100 is a powerful GPU and despite high sparsity in the data features, can still outperform the CPU variant. 

SLIDE is around 1.8 times faster than TF-GPU on Delicious 200k. On the larger Amazon 670k dataset, where we need more computations, the gains are substantially more. SLIDE is around 2.7 (2 hrs vs. 5.5 hrs) times faster than TF-GPU. Most of the computational benefits of SLIDE come from sampling a small subset of active neurons in the output layer. After a few iterations into the training process, the average number of neurons sampled in the output layer for Delicious-200K is $\approx 1000$. Similarly, for Amazon-670K, we sample $\approx 3000$ neurons. With fewer than $0.5\%$ of active neurons, SLIDE outperforms TF-GPU on time by a huge margin on either dataset. It is interesting to note that even after compiling TF-CPU with AVX2 instructions, it is nowhere close to the performance of SLIDE or TF-GPU (SLIDE is 8x faster than TF-CPU). Therefore, it is exciting to note that without any rigorous optimization in our prototype, SLIDE outperforms both baselines using smart randomized algorithms with OpenMP parallelism. 

For Iteration vs. Accuracy plots in Figure \ref{fig:full}, 
we can observe that SLIDE achieves the same accuracy per iteration even though it adaptively selects neurons in some layers. This observation confirms that adaptively selecting neurons and performing asynchronous SGD does not hurt the convergence from an optimization perspective. The plot also confirms that the advantage of SLIDE is not due to any bells and whistles in the optimization process as the convergence with iteration has very similar behavior. For this plot, we only show TF-GPU as the curve for TF-CPU would be identical because the optimization algorithm is the same.  

Since SLIDE performs much fewer computations and memory accesses on the last layer, each iteration is faster than the baselines. This is the critical reason why SLIDE outperforms other baselines when compared on wall-clock time.

\begin{table}
	\centering
		\caption{Core Utilization}
		 \resizebox{0.6\linewidth}{!}{
	\begin{tabular}{|c|c|c|l|} \hline
		& 8&16 & 32\\ \hline
		Tensorflow-CPU & 45$\%$ &35$\%$   & 32$\%$ \\ \hline
		SLIDE & 82$\%$& 81$\%$    & 85$\%$  \\
		\hline\end{tabular}
		}
	\label{table:vtune}
\end{table}

\begin{figure}
    \centering
    \mbox{
        \includegraphics[width=1.5in,height=1.2in]{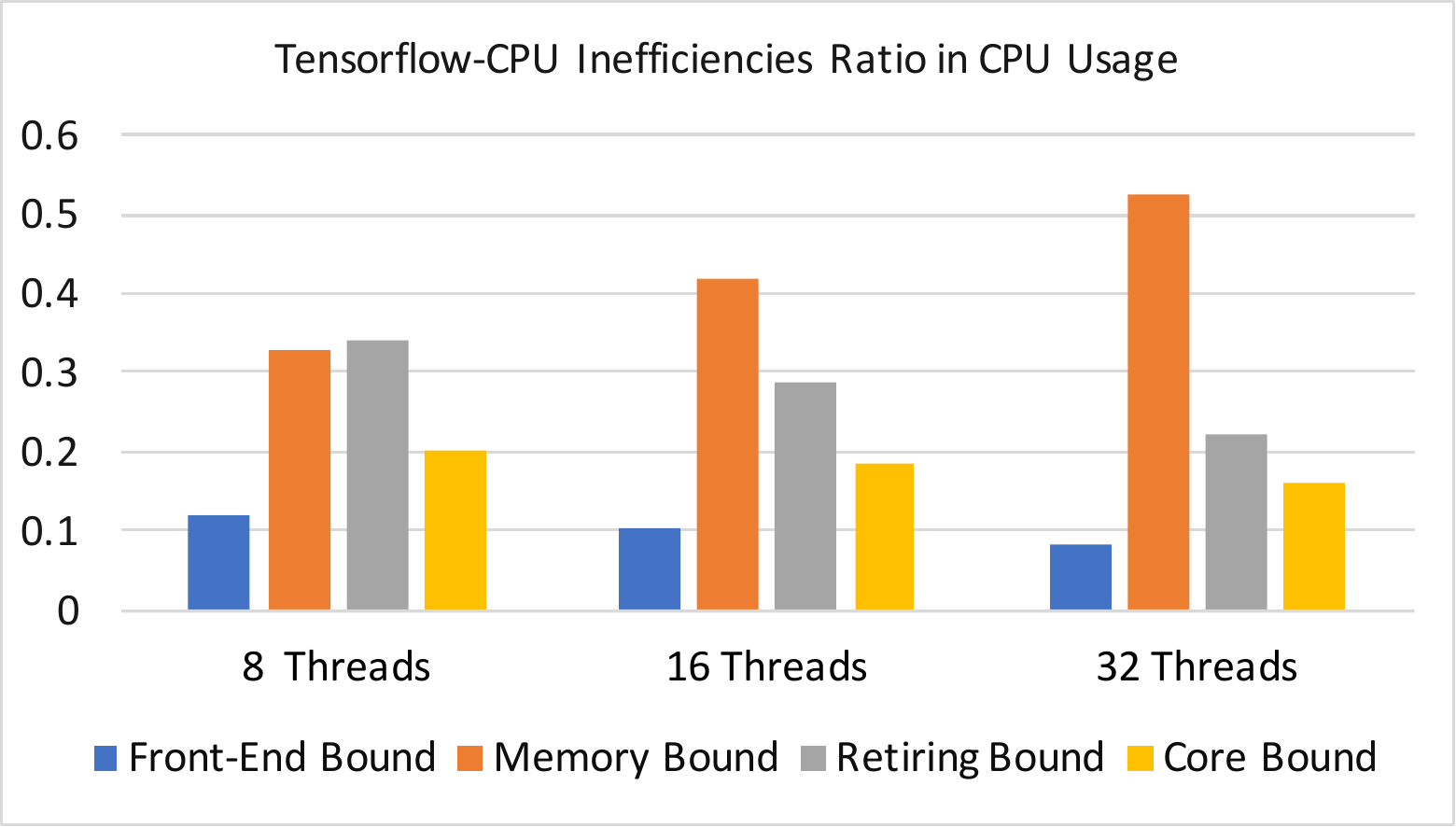}
        \includegraphics[width=1.5in,height=1.2in]{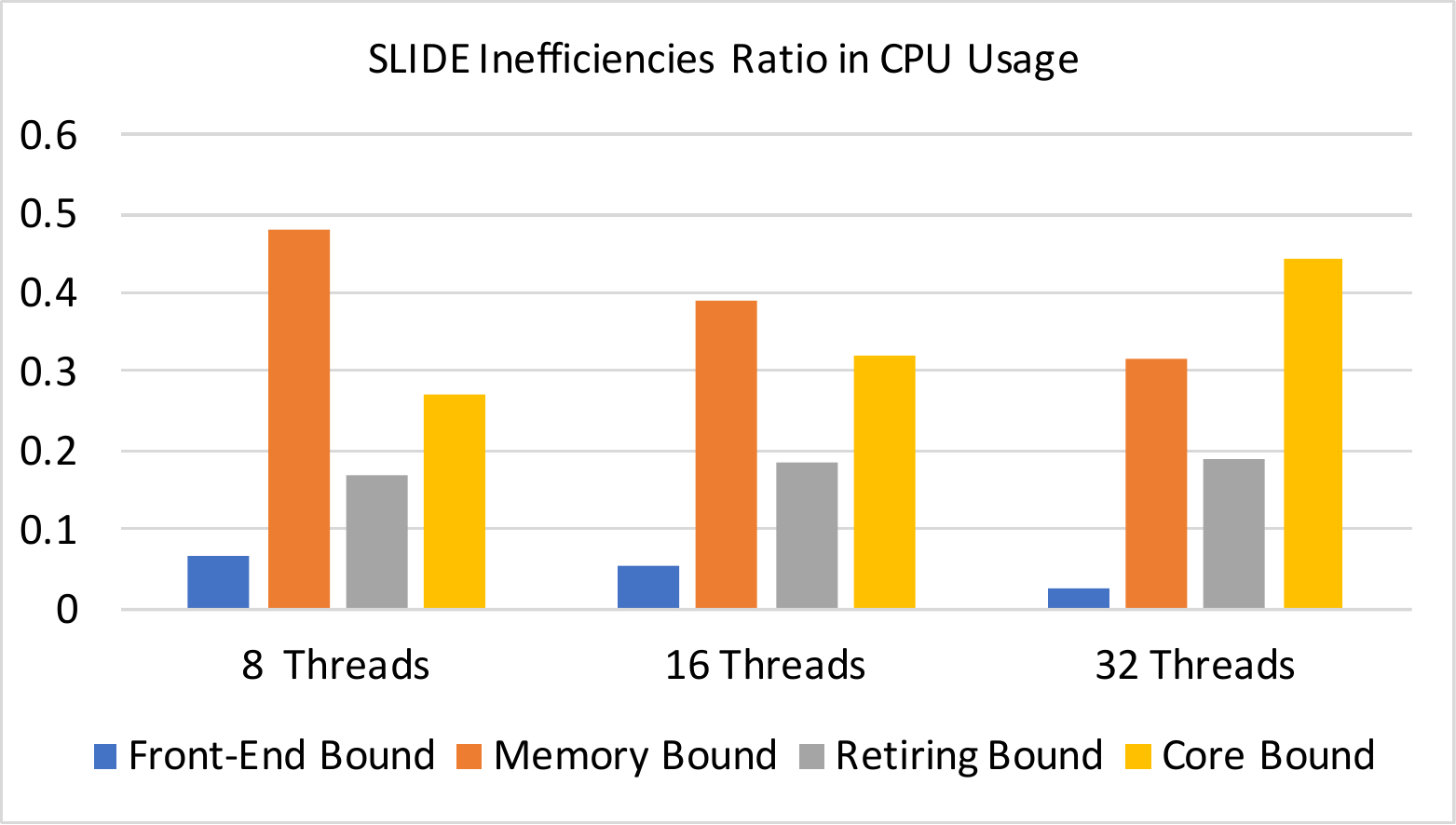}
    }
    \caption{Inefficiencies in CPU Usage: Memory-bound inefficiencies (orange bars) are the most significant ones for either algorithm. For TF-CPU, memory-bound inefficiency rises with an increasing number of cores. For SLIDE, the memory bottleneck reduces with an increasing number of cores. Hence, SLIDE takes better advantage of higher CPU cores, as observed in section~\ref{sec:scalability_tests}.}
    \label{fig:vtune}    
\end{figure}

\textbf{Inefficiency Diagnosis: } We profile and analyze TF-CPU and SLIDE by a state-of-the-art parallel performance analyzer tool, the Intel VTune Performance Analyzer \citep{malladi2009using}. Table~\ref{table:vtune} exhibits the results for core utilization comparison between both frameworks using 8, 16, 32 threads for the above tasks. We can see that for TF-CPU, the utilization is generally low ($<50\%$). It further decreases with more threads. For SLIDE, the core utilization is stable (around $80\%$) across all threads presented in the table~\ref{table:vtune}.

% \hspace{-3cm}
\begin{figure*}[tb]
	\centering
	\mbox{
		\includegraphics[width=1.7in]{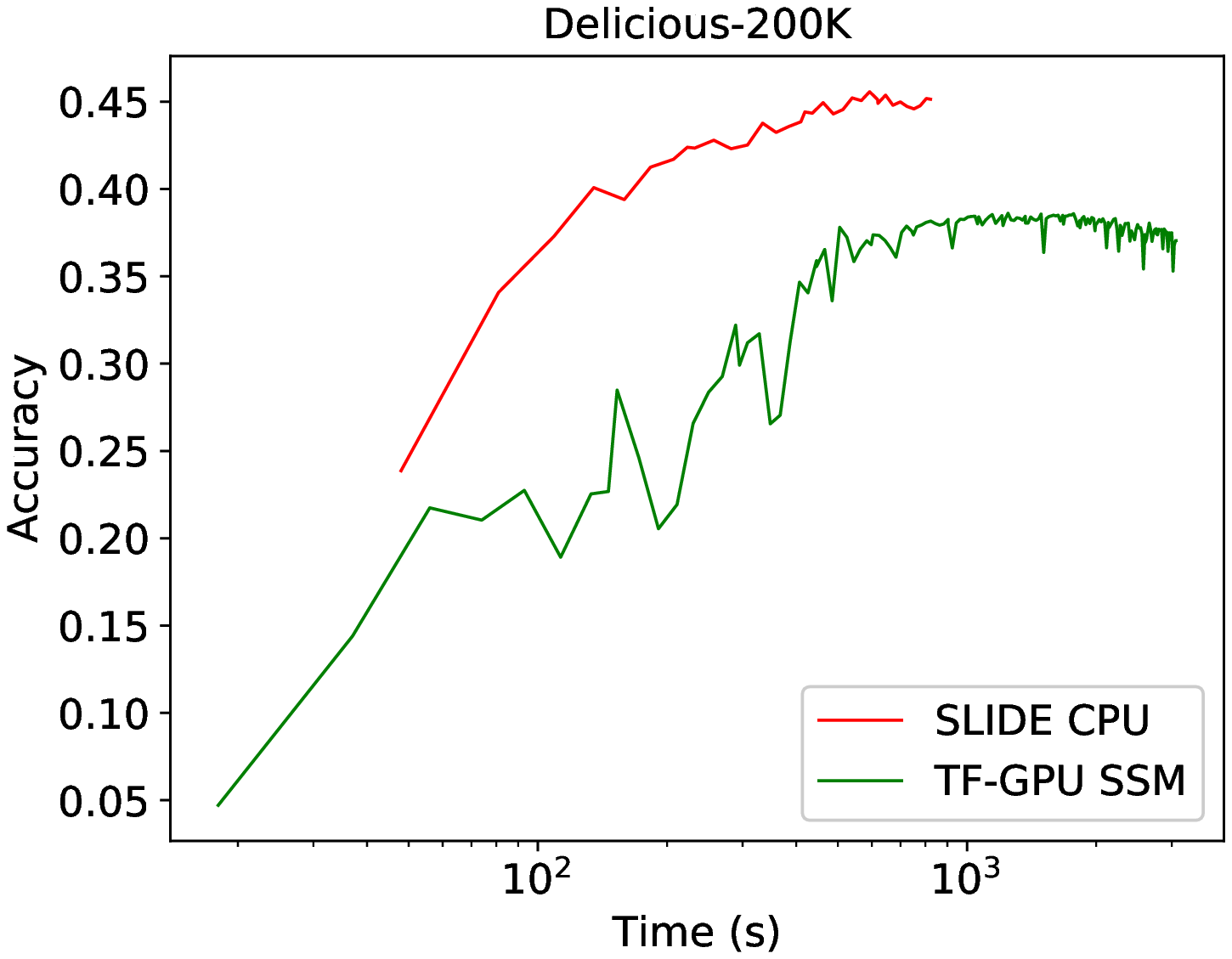}
		\includegraphics[width=1.7in]{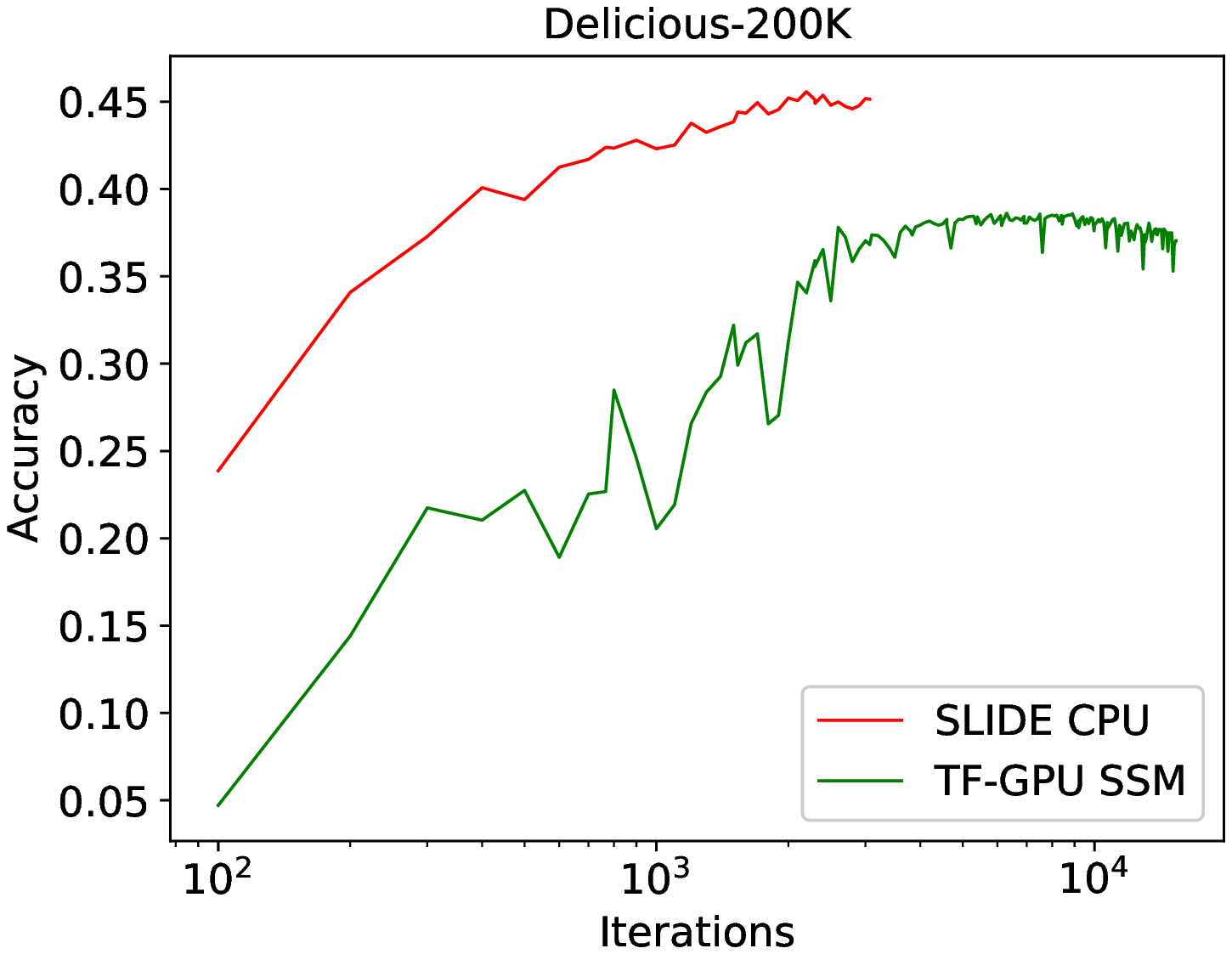}
		\includegraphics[width=1.7in]{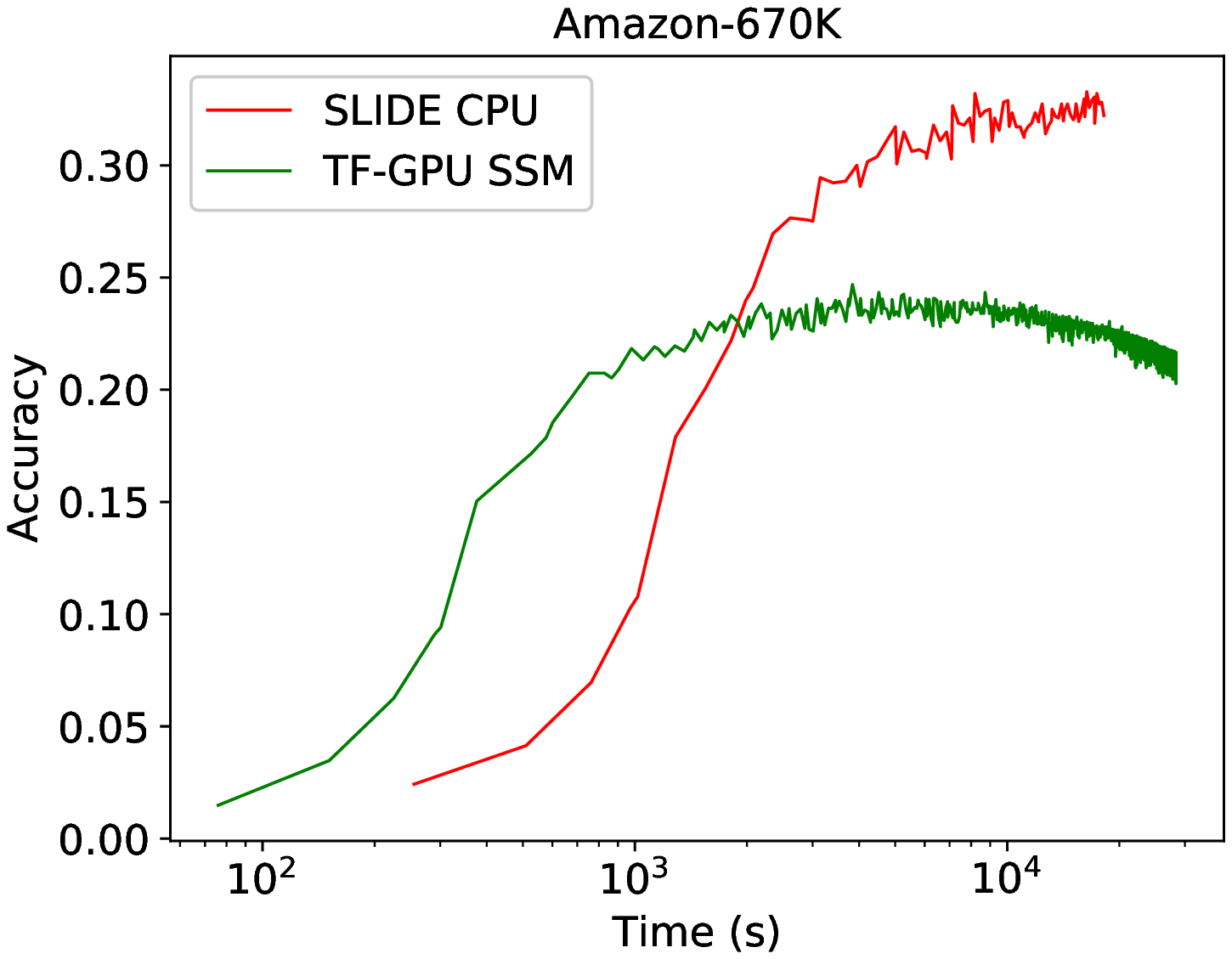}
		\includegraphics[width=1.7in]{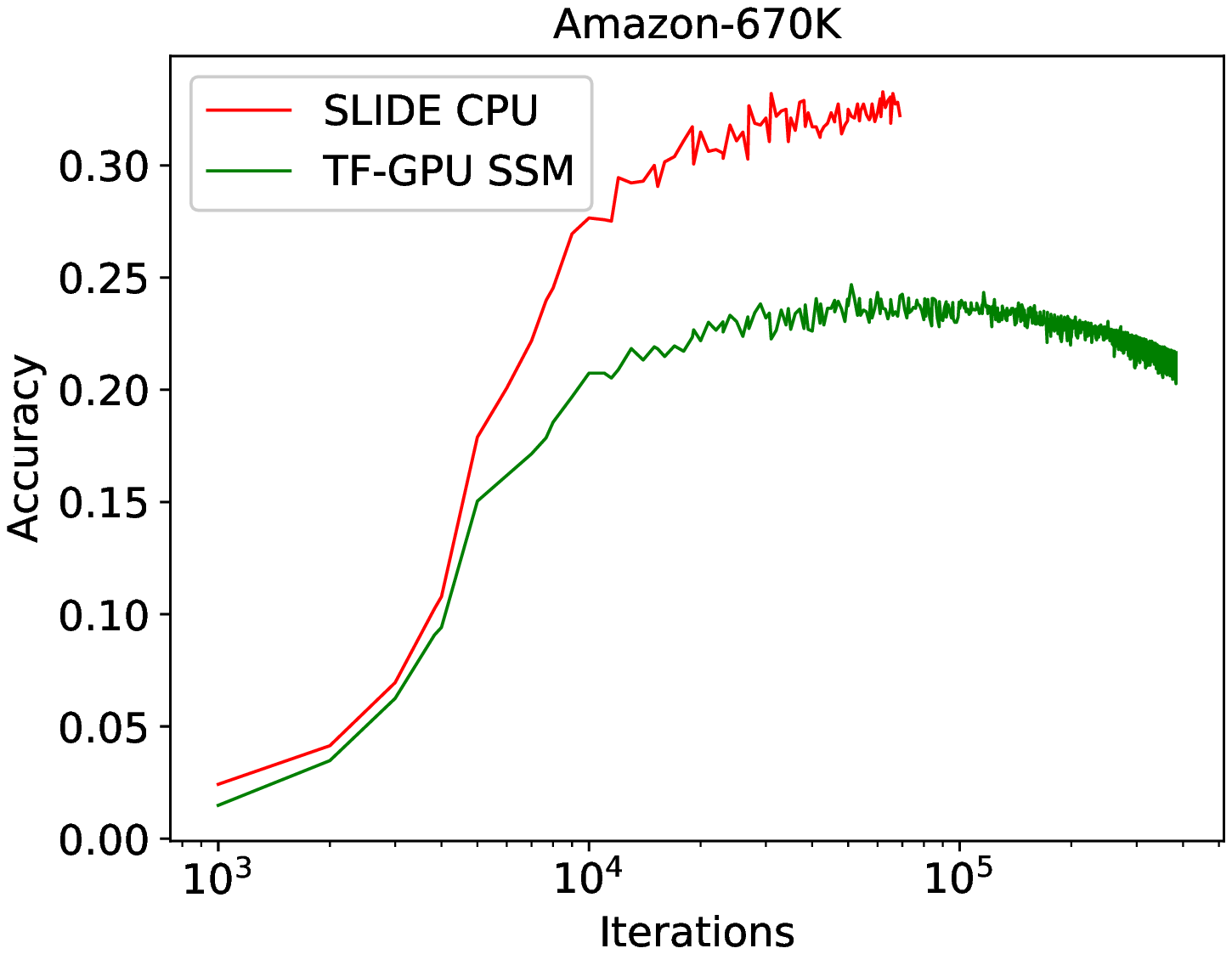}
		}
	    \vspace{-2mm}	
	\caption{It shows the comparison of SLIDE (in red) against the popular Sampled Softmax heuristic (in green). The plots clearly establish the limitations of Sampled Softmax. On Amazon-670K dataset, we notice that Sampled Softmax starts to grow faster than SLIDE in the beginning stages of training but saturates quickly to lower accuracy. SLIDE starts to grow slowly but attains much higher accuracy than Sampled Softmax. SLIDE has the context of choosing the most informative neurons at each layer. Sampled Softmax always chooses a random subset of neurons in the final layer. This reflects in the superior performance of SLIDE over Sampled Softmax.}
	\label{fig:sampled}
\end{figure*}

Figure \ref{fig:vtune} presents the distribution of inefficiencies in CPU usage for TF-CPU and SLIDE. Based on core utilization, the overall inefficiencies of TF-CPU is much more than those of SLIDE. Figure \ref{fig:vtune} provides a detailed distribution of different types of inefficiencies. It is obvious that being memory-bound is a major issue for any number of threads in the histogram. The biggest bottleneck is that a significant fraction of execution pipeline slots are stalled due to demand memory load and store. Observe that the higher the number of cores TF-CPU uses, the more memory-bound it gets.

On the other hand, the higher the number of cores SLIDE uses, the less memory-bound it becomes. Recall that the critical advantage of SLIDE is that it has a lot fewer active neurons and sparse gradient updates. Naturally, memory accesses are a lot fewer than TF-CPU due to very sparse memory accesses within each thread. Our choice of using extra arrays to separate the computations of each thread with asynchronous gradient updates (section~\ref{subsec:design}) across all the threads ensures that simple OpenMP parallelism is sufficient to get near-peak utilization.

\subsection{Comparison with other Heuristics}\label{sec:heuristics}

During the full softmax process in training on Tensorflow, for every training instance, it needs to compute logits (output of the last layer before applying softmax function) for all classes. This step is followed by computing the softmax (normalized sigmoid) of logits. In extreme classification tasks (with a large number of classes), computing these logits gets expensive. Therefore, there has been a line of research working on reducing this cost \citep{mikolov2013distributed, bengio2003quick, gutmann2010noise}. The most common methods are sampling-based (static sampling weights) methods which shortlist a candidate set of classes for every batch of training data. By doing this, the number of computed logits gets reduced significantly. Due to its popularity, Tensorflow supports an optimized implementation of \emph{sampled softmax} \citep{jean2015using}.

We explore how sampled softmax on Tensorflow-GPU performs compared to SLIDE. LSH sampling process in SLIDE is principally very similar to the process of sampled softmax but with sampling probabilities changing dynamically with inputs. We adopt the exact same settings in the previous section for the experiments. Recall that the average number of sampled classes for SLIDE for both the datasets is $\approx 0.5\%$. For sampled softmax, we try a various number of samples. However, with a comparable number of samples, sampled softmax leads to poor accuracy. We empirically observe that we have to sample $20\%$ of the total number of classes to obtain any decent accuracy.

The results are shown in Figure \ref{fig:sampled}. The red lines represent SLIDE, and the green lines represent sampled softmax on Tensorflow-GPU. We can see that both time and iteration wise, the red lines outperform the green lines significantly. Sampled softmax uses static sampling strategies which are fast compared to SLIDE which in contrast uses adaptively changing hash tables for input specific dynamic sampling. Unfortunately, the uninformative static sampling of softmax leads to poor accuracy as shown in the plot. Noted that in these plots, sampled softmax uses significantly more neurons than SLIDE and still shows poor convergence behavior.

Figure \ref{fig:sampled} clearly confirms the need for adaptive sampling of neurons (in proportion to input dependent activation) for sparsifying neural networks in order to retain good convergence. This phenomenon supports our choice of LSH based adaptive sampling.

\subsection{Effect of Batch Size}\label{sec:batch_size}
Batch size is a crucial parameter that can affect the training speed and model quality in Machine Learning. In general, a large batch size may help in reducing the training time per epoch as we process more gradient updates at a time \citep{goyal2017accurate}. But large batches are known to be bad from optimization perspective as they reduce the generalization capability \citep{keskar2016large}. In the case of extreme classification datasets, the number of computations performed is huge owing to large input dimension and a large number of classes. Hence, a larger batch size may not necessarily translate into faster training per epoch. To clarify this, we study the effect of varying batch size on the results. We choose the larger Amazon-670k dataset for this task. Please note that when the batch size is larger than the number of threads, the default scheduling type of OpenMP is static.

\begin{figure*}[t!]
% 	\hspace{-0.25 in}
\centering
		\mbox{
	\includegraphics[width=1.8in]{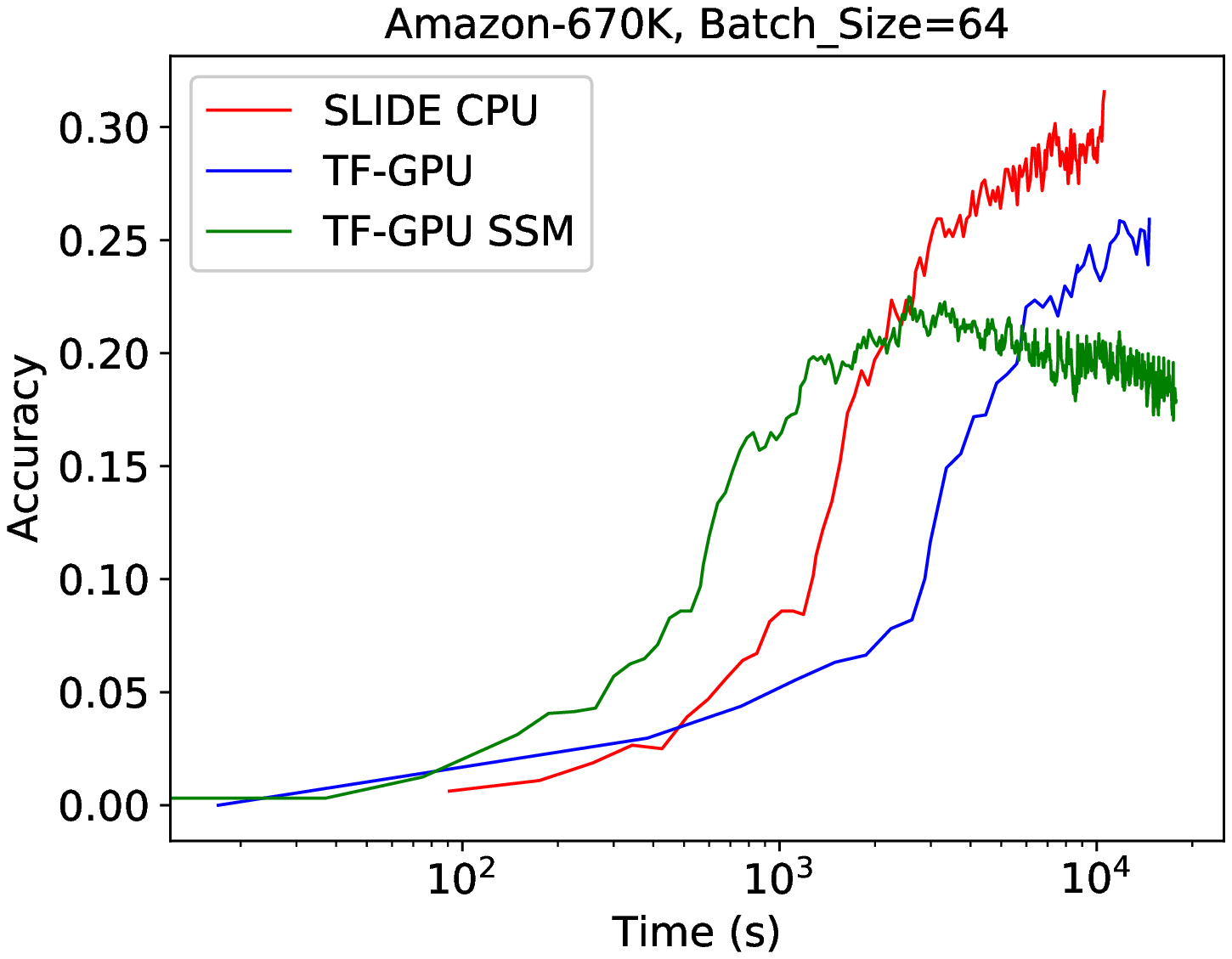}
	\hspace{5mm}
	\includegraphics[width=1.8in]{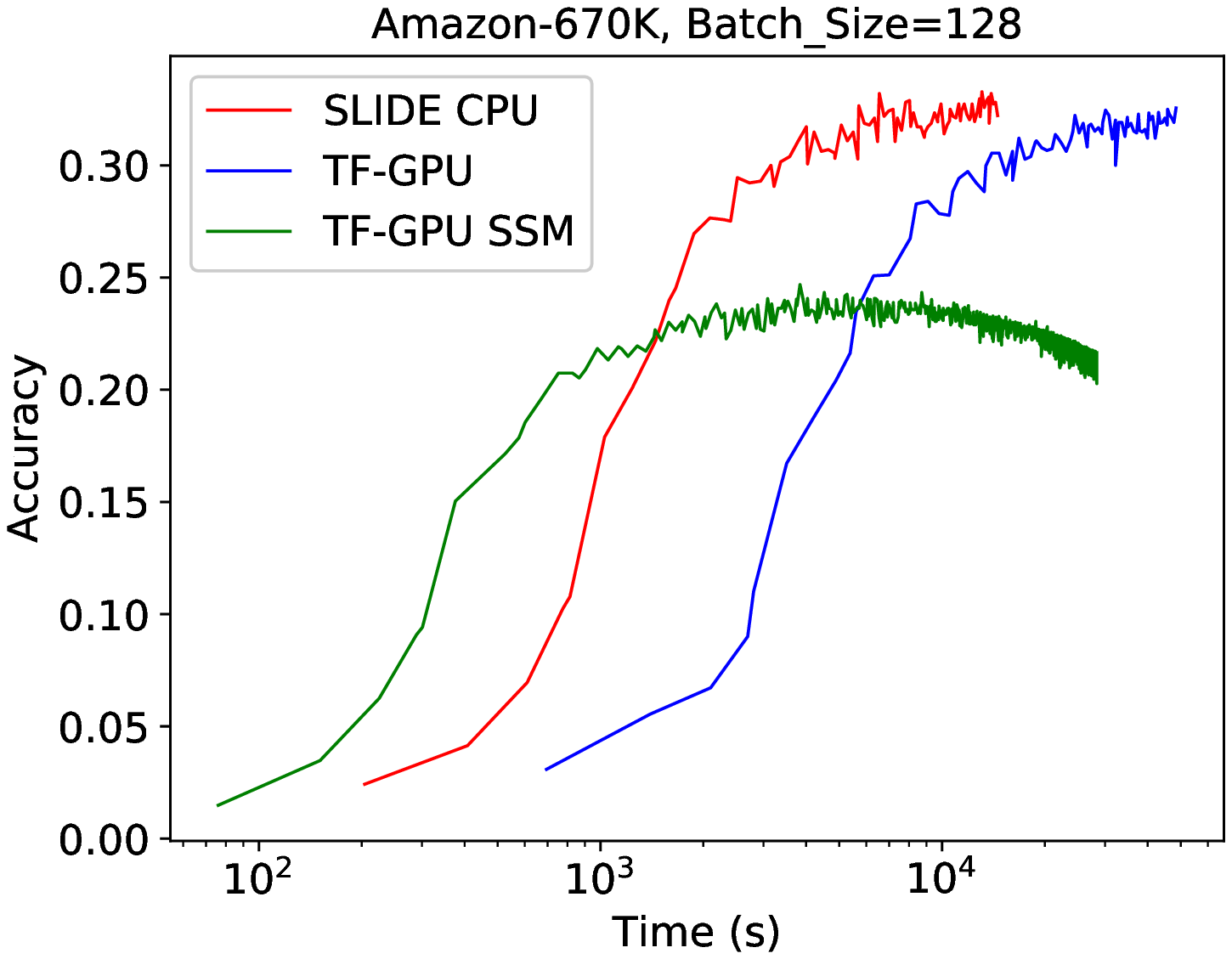}
	\hspace{5mm}
	\includegraphics[width=1.8in]{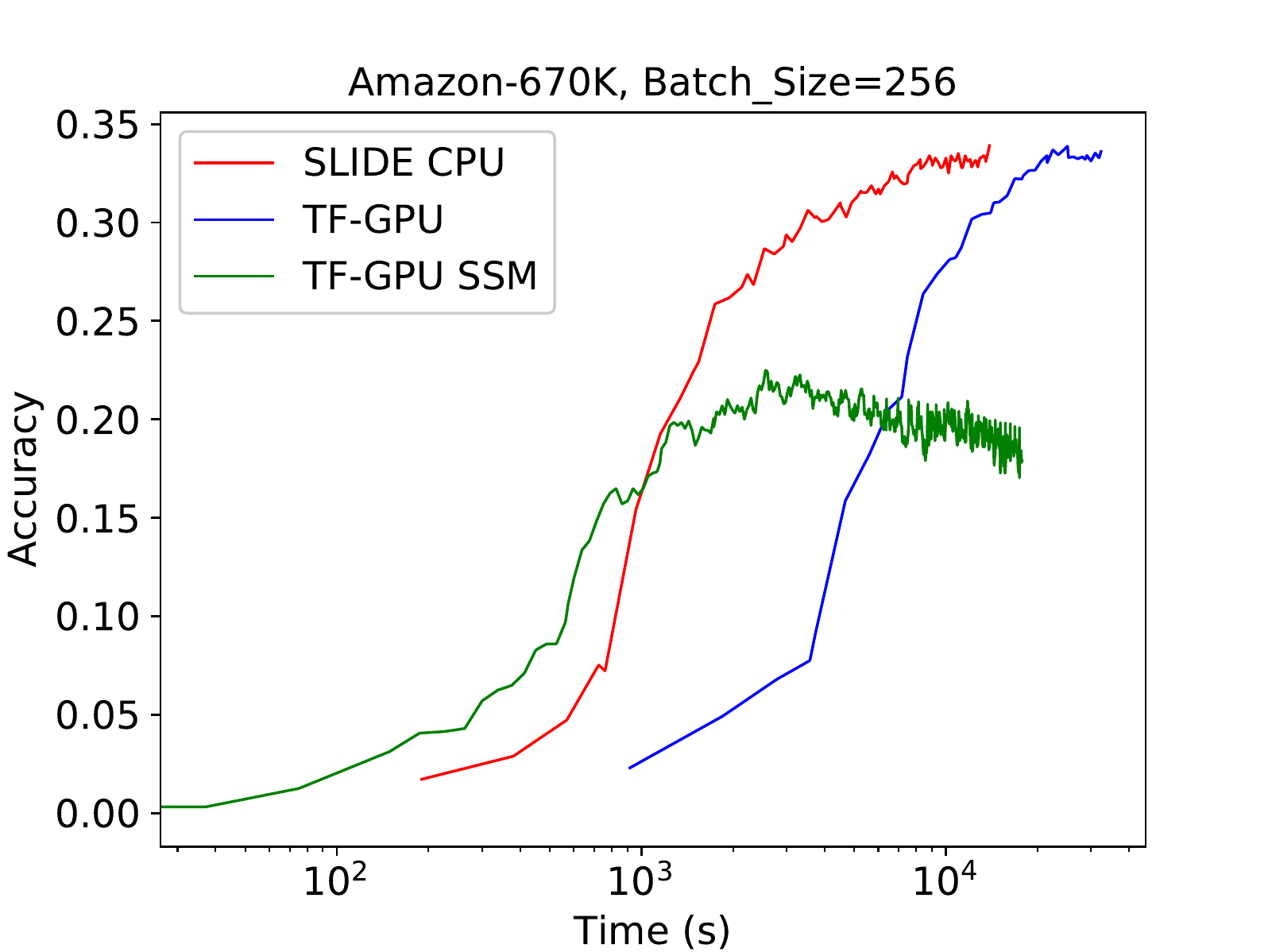}
		}
	\caption{Performance of SLIDE vs. Tensorflow-GPU vs. Sampled Softmax at different batch sizes. SLIDE outperforms the baselines at all batch sizes. As the batch size gets larger, the gap between SLIDE and TF-GPU gets wider.}
	\label{fig:batch}
\end{figure*}

In figure \ref{fig:batch}, we observe that SLIDE outperforms Tensorflow-GPU by a significant margin irrespective of the batch size. This observation could be attributed to the fact that SLIDE performs very few computations per instance. Our data structures allow us to process all samples in a batch in parallel, and the gradient updates are made asynchronously among threads as described in section~\ref{subsec:design}, which enables effective use of parallel threads and it reflects in superior performance over Tensorflow. It is interesting to note that the gap between SLIDE and Tensorflow widens as the batch size grows from 64 to 256.
\subsection{Scalability Tests}\label{sec:scalability_tests}
\begin{figure}[tb]
\vspace{-3mm}
	\centering
	\mbox{
		\includegraphics[width=1.7in]{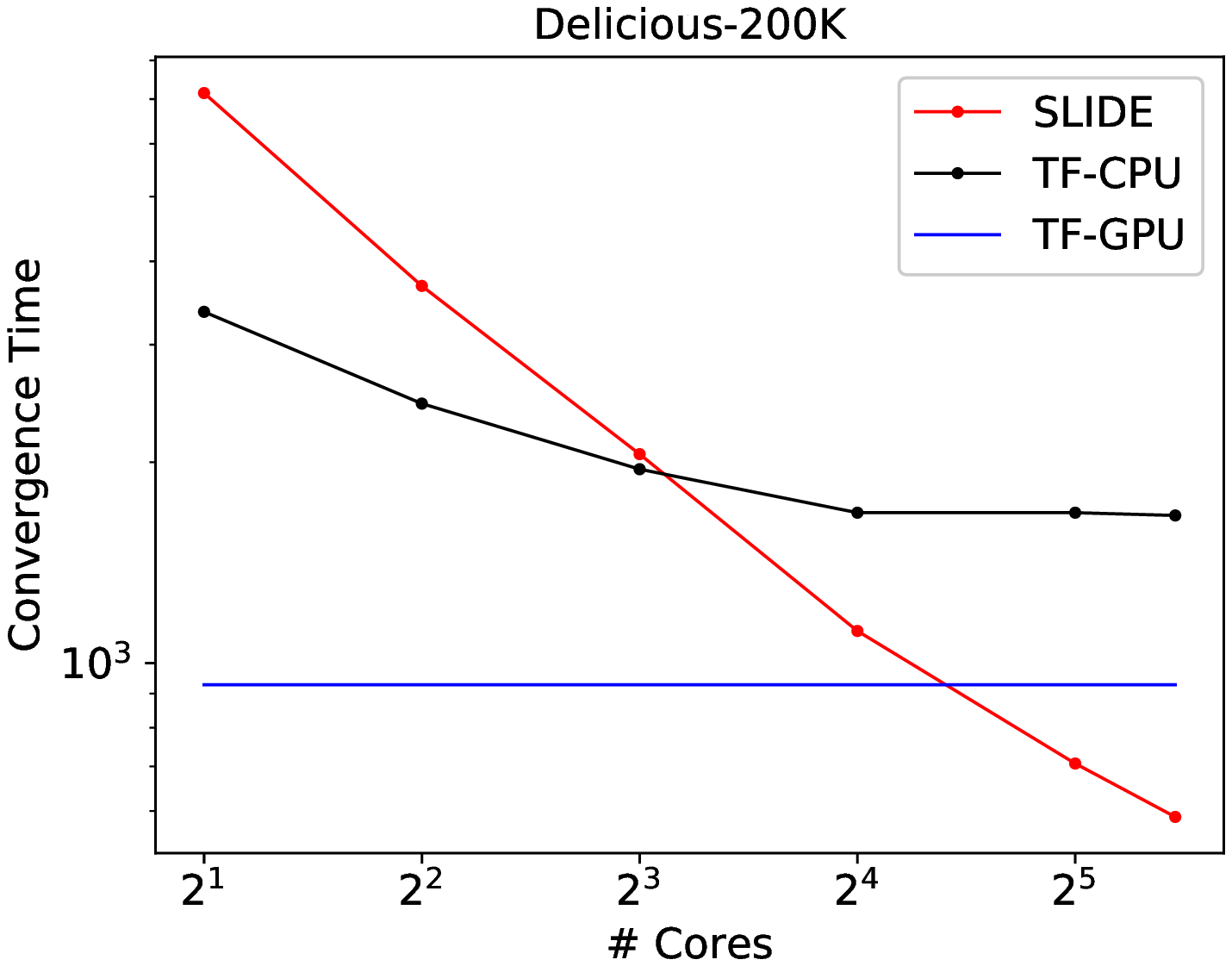}
		\includegraphics[width=1.7in]{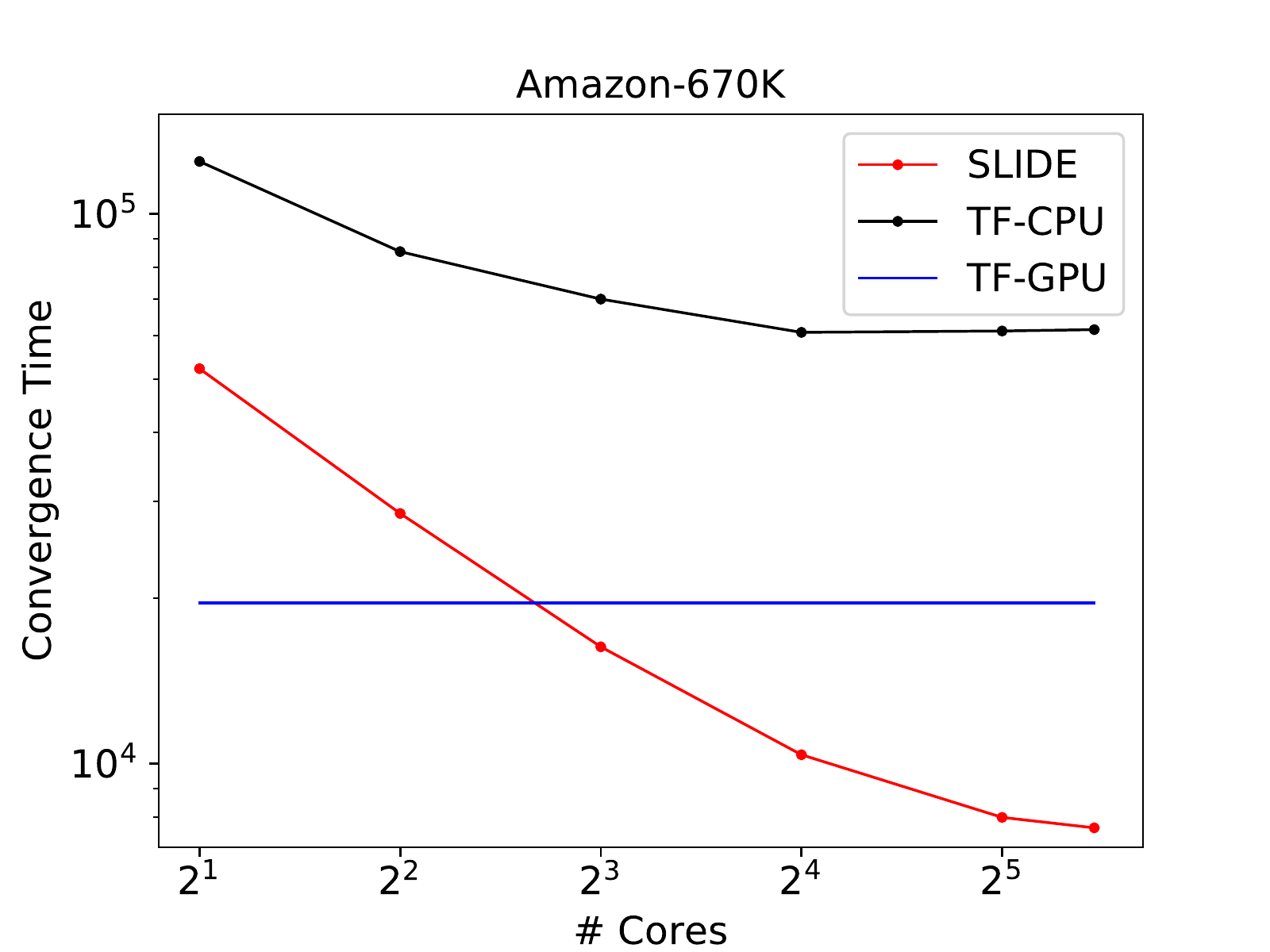}
	}
	    \vspace{-3mm}
	\caption{Scalability Tests: Comparison of performance gains with the number of CPU cores for SLIDE (in red ) vs. Tensorflow-CPU (in black) vs. Tensorflow-GPU (in blue). The blue line is flat because the performance of TF-GPU does not depend on CPU cores. We notice that the convergence time drops steeply for SLIDE compared to TF-CPU/GPU. On Delicious-200K dataset, SLIDE beats TF-CPU with just 8 cores and TF-GPU with less than 32 cores. Similarly, on Amazon-670K dataset, SLIDE beats TF-CPU with just 2 cores and TF-GPU with just 8 cores.}
	\vspace{-3mm}
	\label{fig:scale}
\end{figure}

In this section, we try to understand the effect of increasing CPU cores on the scalability of SLIDE and Tensorflow-CPU. Besides, we intend to know the number of cores SLIDE needs to outperform Tensorflow. As mentioned before, our machine has 44 cores, and each core can have 2 threads. However, we disable multithreading and the effective number of threads and cores is the same. Hence, we interchangeably use the words ``threads" and ``cores" from here on. We benchmark both frameworks with 2, 4, 8, 16, 32, 44 threads.

For the different number of threads, we run the same classification experiments on SLIDE and Tensorflow-CPU for both datasets and clock the corresponding convergence time. Figure \ref{fig:scale} presents the results. The red, blue, black lines represent SLIDE, Tensorflow-GPU, and Tensorflow-CPU, respectively. It should be noted that the blue line is flat because GPU computations were done on V100 with thousands of cores and are mostly oblivious about the number of CPU cores. When the number of cores increases, the convergence time for both SLIDE and Tensorflow-CPU starts to decrease. This decrease is expected due to the benefits brought by more parallelism on each training batch. For Delicious dataset, the red line and the black line cross each other at around 8 cores, which means that with more than 8 cores, SLIDE can beat Tensorflow-CPU. The red and blue lines intersect between 16 and 32 cores. Hence, with fewer than 32 cores, SLIDE outperforms Tensorflow-GPU on Delicious dataset. Similarly, for larger Amazon dataset, the red and black line never intersect, and the red and blue line intersects on 8 cores. This means that SLIDE beats Tensorflow-GPU with as few as 8 CPU cores and Tensorflow-CPU with as few as 2 CPU cores.

% Moreover, based on the statistics collected through experiments as mentioned above, we show the ratio of convergence time with the different number of cores to the minimum convergence time (using 44 cores). The results are exhibited in Figure \ref{fig:scale}. Again, the red line represents SLIDE, and the black line represents Tensorflow-CPU. When the number of cores increases, that ratio decreases for both SLIDE and Tensorflow-CPU. However, it is explicit that the ratio drops more drastically for the red line than the black line. This behavior concludes that the scalability of SLIDE is much better than that of Tensorflow-CPU. Moreover, in the plot, we observe that the benefits of using more cores are not obvious after 16 cores for Tensorflow-CPU. Coincidentally, a very recent work \citep{DBLP:journals/corr/abs-1812-01665} introduces the hardness of finding the optimal parameter settings of Tensorflow’s threading model for CPU backends. It argues that getting the best performance from a CPU needs manual, tedious and time-consuming tuning and it still may not guarantee the best performance. While analyzing the scalability and core utilization of Tensorflow-CPU can be an independent research interest, we explore a small aspect of it in the following paragraphs.

\subsection{Additional Speedup with Threading Model and Platform Micro-architecture}\label{sec:Hugepages}
\begin{figure}[t]
    \centering
    \vspace{-3mm}
    \includegraphics[width=1.7in]{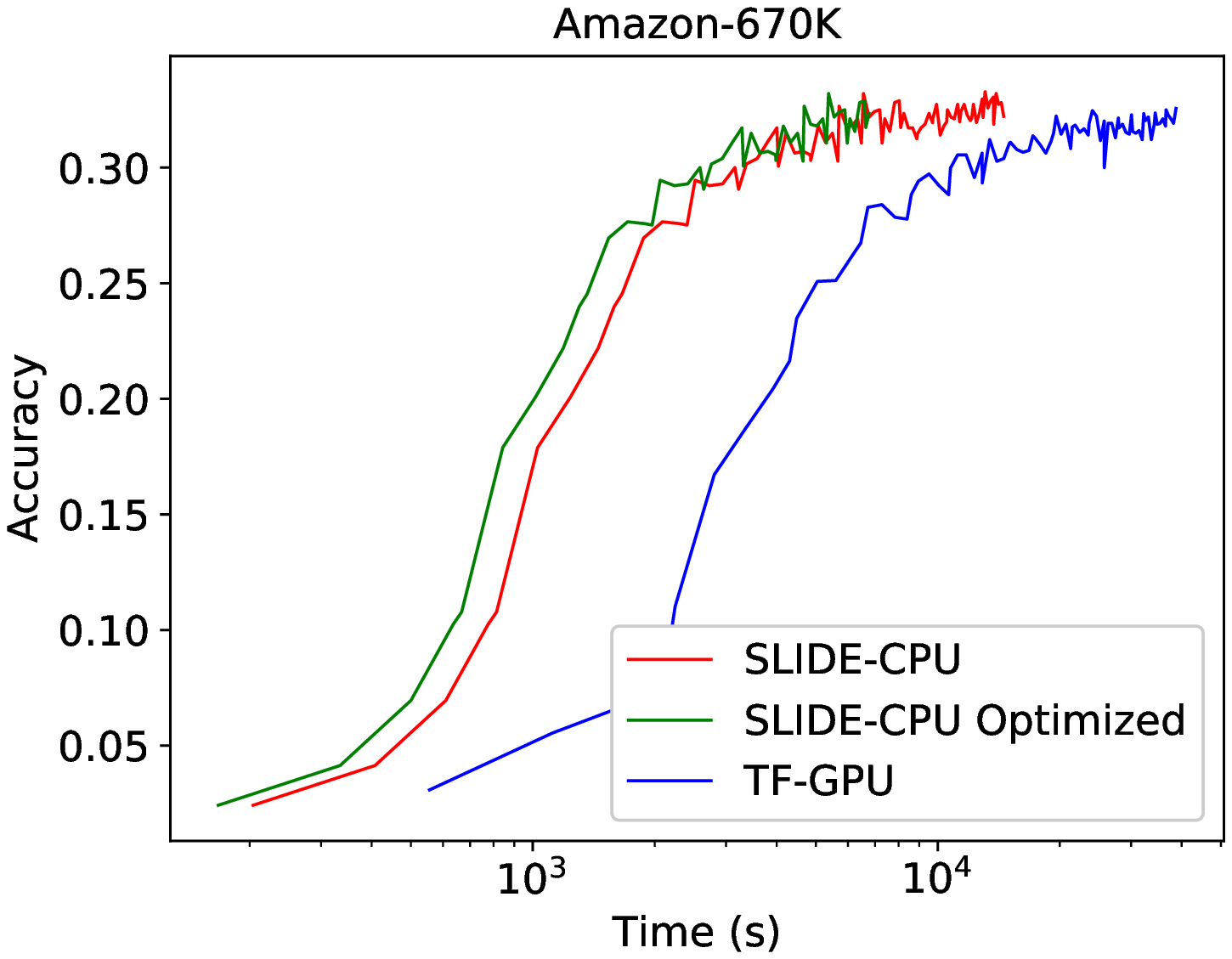}
    \hspace{-6mm}
    \includegraphics[width=1.7in]{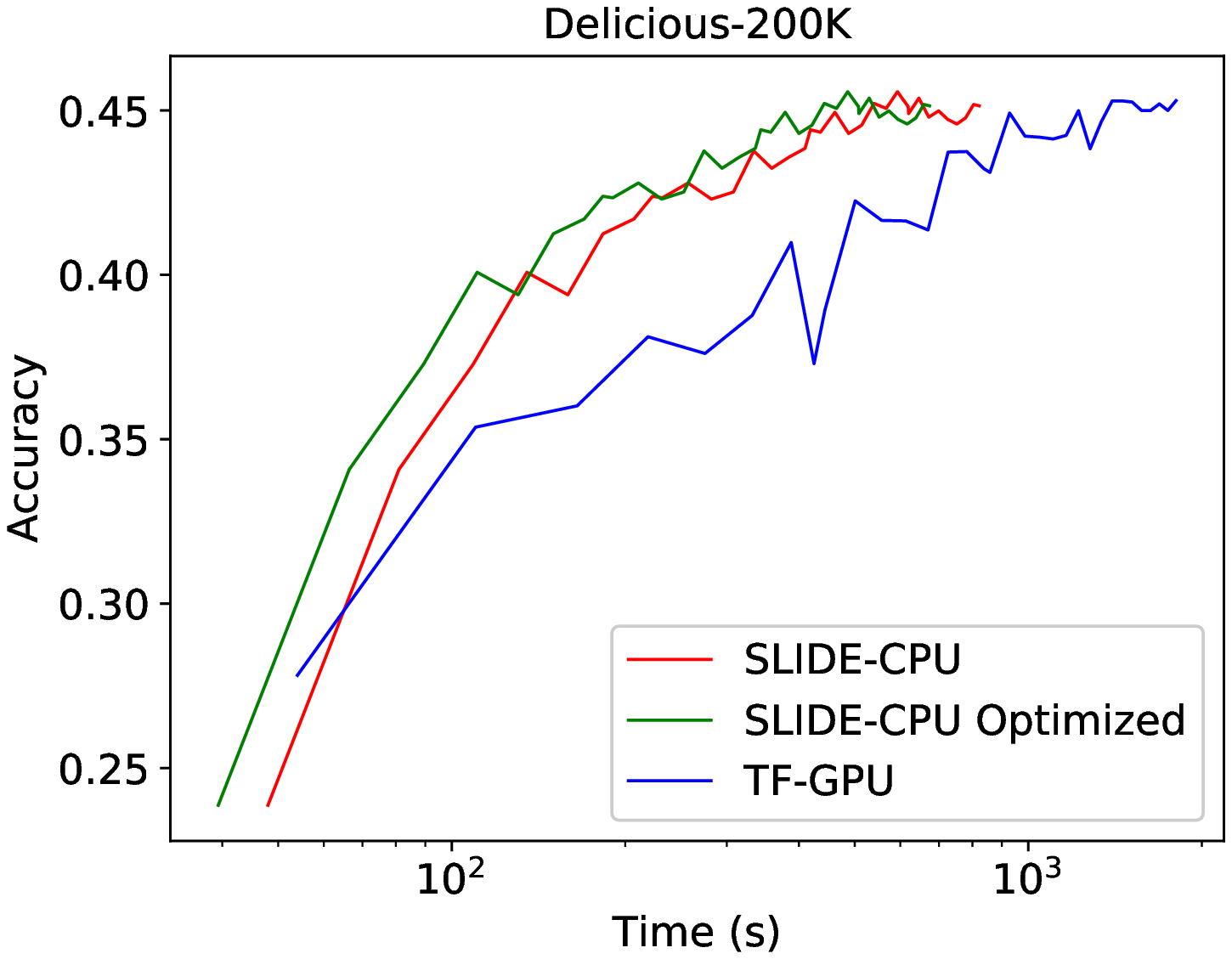}
    \vspace{-3mm}
    \caption{Impact of Hugepages and SIMD Optimization: The comparison of training time for optimized version of SLIDE against a plain version of SLIDE and TF-GPU. We can see that SLIDE-Optimized is roughly 1.3x faster than the un-optimized one on both datasets (x-axis is log scale).}
        \label{fig:Hugepages}
        \vspace{-3mm}
\end{figure}

In this section, we perform several CPU optimizations outlined in appendix \ref{sec:HPC} to reduce cache misses. We first install $Hugepages$ package for Ubuntu, which offers 2MB and 1GB cache pages in addition to default 4KB ones. We pre-allocate 1000 2MB Hugepages and 10 1GB Hugepages which are found to be enough for both Delicious-200K and Amazon-670K datasets. To resolve the issue of the false sharing for OpenMP mutli-thread, we give a provision to our data structures to align on cache line boundaries. Besides using Hugepages, we also used SIMD instructions (specifically, Intel-AVX) to facilitate per thread batching. In figure \ref{fig:Hugepages}, we compare the benefit of aforementioned optimizations against an un-optimized SLIDE and TF-GPU. We notice that Cache-Optimized SLIDE (in green) is $\approx 1.3$ times faster than basic SLIDE (in red). Since we already have a 2.7x speed-up over TF-GPU on Amazon-670K, it translates to 3.5x speedup over TF-GPU and a 10x speedup over TF-CPU.

In appendix \ref{sec:tlb}, we measure the impact of HugePages on various CPU-counter metrics like {\it TLB} miss rates and {\it PageFaults}. Concisely, we notice that {\it HugePages} reduces the misses by a large margin.
% \begin{wrapfigure}{R}{5cm}
% \end{wrapfigure}

\section{Conclusion and Future Work}

We provide the first evidence that a smart algorithm with modest CPU OpenMP parallelism can outperform the best available hardware NVIDIA-V100, for training large deep learning architectures. Our system SLIDE is a combination of carefully tailored randomized hashing algorithms with the right data structures that allow asynchronous parallelism. We show up to 3.5x gain against TF-GPU and 10x gain against TF-CPU in training time with similar precision on popular extreme classification datasets. Our next steps are to extend SLIDE to include convolutional layers. SLIDE has unique benefits when it comes to random memory accesses and parallelism. We anticipate that a distributed implementation of SLIDE would be very appealing because the communication costs are minimal due to sparse gradients. 

\section{Acknowledgements}
The work was supported by NSF-1652131, NSF-BIGDATA 1838177, AFOSR-YIPFA9550-18-1-0152, Amazon Research Award, and ONR BRC grant for Randomized Numerical Linear Algebra.

\bibliography{main}

\begin{thebibliography}{40}
\providecommand{\natexlab}[1]{#1}
\providecommand{\url}[1]{\texttt{#1}}
\expandafter\ifx\csname urlstyle\endcsname\relax
  \providecommand{\doi}[1]{doi: #1}\else
  \providecommand{\doi}{doi: \begingroup \urlstyle{rm}\Url}\fi

\bibitem[Ba \& Frey(2013)Ba and Frey]{ba2013adaptive}
Ba, J. and Frey, B.
\newblock Adaptive dropout for training deep neural networks.
\newblock In \emph{Advances in Neural Information Processing Systems}, pp.\
  3084--3092, 2013.

\bibitem[Basu et~al.(2013)Basu, Gandhi, Chang, Hill, and Swift]{basu2013paging}
Basu, A., Gandhi, J., Chang, J., Hill, M., and Swift, M.
\newblock Efficient virtual memory for big memory servers.
\newblock In \emph{International Symposium on Computer Architecture}, pp.\
  237--248, 2013.

\bibitem[Bengio et~al.()]{bengio2003quick}
Bengio, Y. et~al.
\newblock Quick training of probabilistic neural nets by importance sampling.

\bibitem[Blanc \& Rendle(2018)Blanc and Rendle]{blanc2018adaptive}
Blanc, G. and Rendle, S.
\newblock Adaptive sampled softmax with kernel based sampling.
\newblock In \emph{International Conference on Machine Learning}, pp.\
  589--598, 2018.

\bibitem[Chen \& Shrivastava(2018)Chen and Shrivastava]{chen2018densified}
Chen, B. and Shrivastava, A.
\newblock Densified winner take all (wta) hashing for sparse datasets.
\newblock In \emph{Uncertainty in artificial intelligence}, 2018.

\bibitem[Chen et~al.(2018)Chen, Shrivastava, and Steorts]{chen2018unique}
Chen, B., Shrivastava, A., and Steorts, R.~C.
\newblock Unique entity estimation with application to the syrian conflict.
\newblock \emph{THE ANNALS}, 2018.

\bibitem[Chen et~al.(2019)Chen, Xu, and Shrivastava]{chen2019fast}
Chen, B., Xu, Y., and Shrivastava, A.
\newblock Fast and accurate stochastic gradient estimation.
\newblock In \emph{Advances in Neural Information Processing Systems}, pp.\
  12339--12349, 2019.

\bibitem[Corbet(2011)]{Corbet2011THP}
Corbet, J.
\newblock Transparent huge pages in 2.6.38.
\newblock \emph{http://lwn.net/Articles/423584/}, 2011.

\bibitem[Gionis et~al.(1999)Gionis, Indyk, and Motwani]{gionis1999similarity}
Gionis, A., Indyk, P., and Motwani, R.
\newblock Similarity search in high dimensions via hashing.
\newblock In \emph{Proceedings of the 25th International Conference on Very
  Large Data Bases}, VLDB '99, pp.\  518--529, San Francisco, CA, USA, 1999.
  Morgan Kaufmann Publishers Inc.
\newblock ISBN 1-55860-615-7.
\newblock URL \url{http://dl.acm.org/citation.cfm?id=645925.671516}.

\bibitem[Goyal et~al.(2017)Goyal, Doll{\'a}r, Girshick, Noordhuis, Wesolowski,
  Kyrola, Tulloch, Jia, and He]{goyal2017accurate}
Goyal, P., Doll{\'a}r, P., Girshick, R., Noordhuis, P., Wesolowski, L., Kyrola,
  A., Tulloch, A., Jia, Y., and He, K.
\newblock Accurate, large minibatch sgd: Training imagenet in 1 hour.
\newblock \emph{arXiv preprint arXiv:1706.02677}, 2017.

\bibitem[Gutmann \& Hyv{\"a}rinen(2010)Gutmann and
  Hyv{\"a}rinen]{gutmann2010noise}
Gutmann, M. and Hyv{\"a}rinen, A.
\newblock Noise-contrastive estimation: A new estimation principle for
  unnormalized statistical models.
\newblock In \emph{Proceedings of the Thirteenth International Conference on
  Artificial Intelligence and Statistics}, pp.\  297--304, 2010.

\bibitem[Hasabnis(2018)]{DBLP:journals/corr/abs-1812-01665}
Hasabnis, N.
\newblock Auto-tuning tensorflow threading model for {CPU} backend.
\newblock \emph{CoRR}, abs/1812.01665, 2018.
\newblock URL \url{http://arxiv.org/abs/1812.01665}.

\bibitem[Indyk \& Motwani(1998)Indyk and Motwani]{indyk1998approximate}
Indyk, P. and Motwani, R.
\newblock Approximate nearest neighbors: towards removing the curse of
  dimensionality.
\newblock In \emph{Proceedings of the thirtieth annual ACM symposium on Theory
  of computing}, pp.\  604--613. ACM, 1998.

\bibitem[Jean et~al.(2015)Jean, Cho, Memisevic, and Bengio]{jean2015using}
Jean, S., Cho, K., Memisevic, R., and Bengio, Y.
\newblock On using very large target vocabulary for neural machine translation.
\newblock In \emph{Proceedings of the 53rd Annual Meeting of the Association
  for Computational Linguistics and the 7th International Joint Conference on
  Natural Language Processing (Volume 1: Long Papers)}, volume~1, pp.\  1--10,
  2015.

\bibitem[Jouppi et~al.(2017)Jouppi, Young, Patil, Patterson, Agrawal, Bajwa,
  Bates, Bhatia, Boden, Borchers, et~al.]{jouppi2017datacenter}
Jouppi, N.~P., Young, C., Patil, N., Patterson, D., Agrawal, G., Bajwa, R.,
  Bates, S., Bhatia, S., Boden, N., Borchers, A., et~al.
\newblock In-datacenter performance analysis of a tensor processing unit.
\newblock In \emph{2017 ACM/IEEE 44th Annual International Symposium on
  Computer Architecture (ISCA)}, pp.\  1--12. IEEE, 2017.

\bibitem[Karakostas et~al.(2014)Karakostas, Unsal, Nemirovsky, Cristal, and
  Swift]{karakos2014paging}
Karakostas, V., Unsal, O., Nemirovsky, M., Cristal, A., and Swift, M.
\newblock Performance analysis of the memory management unit under scale-out
  workloads.
\newblock In \emph{International Symposium on Workload Characterization}, pp.\
  1--12, 2014.

\bibitem[Keskar et~al.(2016)Keskar, Mudigere, Nocedal, Smelyanskiy, and
  Tang]{keskar2016large}
Keskar, N.~S., Mudigere, D., Nocedal, J., Smelyanskiy, M., and Tang, P. T.~P.
\newblock On large-batch training for deep learning: Generalization gap and
  sharp minima.
\newblock \emph{arXiv preprint arXiv:1609.04836}, 2016.

\bibitem[Kingma \& Ba(2014)Kingma and Ba]{kingma2014adam}
Kingma, D.~P. and Ba, J.
\newblock Adam: A method for stochastic optimization.
\newblock \emph{arXiv preprint arXiv:1412.6980}, 2014.

\bibitem[Kumar et~al.(2017)Kumar, Soltis, Esmer, Yoaz, and Kottapalli]{skylake}
Kumar, A., Soltis, D., Esmer, I., Yoaz, I., and Kottapalli, S.
\newblock The new intel xeon scalable processor(formerly skylake-sp).
\newblock In \emph{Hot Chips}, 2017.

\bibitem[Kush~Bhatia()]{extreme}
Kush~Bhatia, Kunal~Dahiya, H. J. Y. P. M.~V.
\newblock The extreme classification repository: Multi-label datasets and code.
\newblock \url{http://manikvarma.org/downloads/XC/XMLRepository.html#Prabhu14}.

\bibitem[Le~Gall(2014)]{le2014powers}
Le~Gall, F.
\newblock Powers of tensors and fast matrix multiplication.
\newblock In \emph{Proceedings of the 39th international symposium on symbolic
  and algebraic computation}, pp.\  296--303. ACM, 2014.

\bibitem[Li et~al.(2006)Li, Hastie, and Church]{li2006very}
Li, P., Hastie, T.~J., and Church, K.~W.
\newblock Very sparse random projections.
\newblock In \emph{Proceedings of the 12th ACM SIGKDD international conference
  on Knowledge discovery and data mining}, pp.\  287--296. ACM, 2006.

\bibitem[Luo \& Shrivastava(2018)Luo and Shrivastava]{luo2018scaling}
Luo, C. and Shrivastava, A.
\newblock Scaling-up split-merge mcmc with locality sensitive sampling (lss).
\newblock \emph{arXiv preprint arXiv:1802.07444}, 2018.

\bibitem[Makhzani \& Frey(2013)Makhzani and Frey]{makhzani2013k}
Makhzani, A. and Frey, B.
\newblock K-sparse autoencoders.
\newblock \emph{arXiv preprint arXiv:1312.5663}, 2013.

\bibitem[Makhzani \& Frey(2015)Makhzani and Frey]{makhzani2015winner}
Makhzani, A. and Frey, B.~J.
\newblock Winner-take-all autoencoders.
\newblock In \emph{Advances in neural information processing systems}, pp.\
  2791--2799, 2015.

\bibitem[Malladi(2009)]{malladi2009using}
Malladi, R.~K.
\newblock Using intel{\textregistered} vtune™ performance analyzer
  events/ratios \& optimizing applications.
\newblock \emph{http:/software. intel. com}, 2009.

\bibitem[Meng \& Skadron(2009)Meng and Skadron]{meng09llc}
Meng, J. and Skadron, K.
\newblock Avoiding cache thrashing due to private data placement in last-level
  cache for manycore scaling.
\newblock In \emph{International Conference on Computer Design}, pp.\
  283--297, 2009.

\bibitem[Mikolov et~al.(2013)Mikolov, Sutskever, Chen, Corrado, and
  Dean]{mikolov2013distributed}
Mikolov, T., Sutskever, I., Chen, K., Corrado, G.~S., and Dean, J.
\newblock Distributed representations of words and phrases and their
  compositionality.
\newblock In \emph{Advances in neural information processing systems}, pp.\
  3111--3119, 2013.

\bibitem[Owens et~al.(2008)Owens, Houston, Luebke, Green, Stone, and
  Phillips]{owens2008gpu}
Owens, J.~D., Houston, M., Luebke, D., Green, S., Stone, J.~E., and Phillips,
  J.~C.
\newblock Gpu computing.
\newblock 2008.

\bibitem[Recht et~al.(2011)Recht, Re, Wright, and Niu]{recht2011hogwild}
Recht, B., Re, C., Wright, S., and Niu, F.
\newblock Hogwild: A lock-free approach to parallelizing stochastic gradient
  descent.
\newblock In \emph{Advances in neural information processing systems}, pp.\
  693--701, 2011.

\bibitem[Shrivastava \& Li(2014{\natexlab{a}})Shrivastava and
  Li]{shrivastava2014asymmetric}
Shrivastava, A. and Li, P.
\newblock Asymmetric lsh (alsh) for sublinear time maximum inner product search
  (mips).
\newblock In \emph{Advances in Neural Information Processing Systems}, pp.\
  2321--2329, 2014{\natexlab{a}}.

\bibitem[Shrivastava \& Li(2014{\natexlab{b}})Shrivastava and
  Li]{shrivastava2014densifying}
Shrivastava, A. and Li, P.
\newblock Densifying one permutation hashing via rotation for fast near
  neighbor search.
\newblock In \emph{International Conference on Machine Learning}, pp.\
  557--565, 2014{\natexlab{b}}.

\bibitem[Spring \& Shrivastava(2017{\natexlab{a}})Spring and
  Shrivastava]{spring2017new}
Spring, R. and Shrivastava, A.
\newblock A new unbiased and efficient class of lsh-based samplers and
  estimators for partition function computation in log-linear models.
\newblock \emph{arXiv preprint arXiv:1703.05160}, 2017{\natexlab{a}}.

\bibitem[Spring \& Shrivastava(2017{\natexlab{b}})Spring and
  Shrivastava]{spring2017scalable}
Spring, R. and Shrivastava, A.
\newblock Scalable and sustainable deep learning via randomized hashing.
\newblock In \emph{Proceedings of the 23rd ACM SIGKDD International Conference
  on Knowledge Discovery and Data Mining}, pp.\  445--454. ACM,
  2017{\natexlab{b}}.

\bibitem[Srivastava et~al.(2014)Srivastava, Hinton, Krizhevsky, Sutskever, and
  Salakhutdinov]{srivastava2014dropout}
Srivastava, N., Hinton, G., Krizhevsky, A., Sutskever, I., and Salakhutdinov,
  R.
\newblock Dropout: a simple way to prevent neural networks from overfitting.
\newblock \emph{The Journal of Machine Learning Research}, 15\penalty0
  (1):\penalty0 1929--1958, 2014.

\bibitem[Vitter(1985)]{vitter1985random}
Vitter, J.~S.
\newblock Random sampling with a reservoir.
\newblock \emph{ACM Transactions on Mathematical Software (TOMS)}, 11\penalty0
  (1):\penalty0 37--57, 1985.

\bibitem[Wang et~al.(2018)Wang, Shrivastava, Wang, and Ryu]{wang2018randomized}
Wang, Y., Shrivastava, A., Wang, J., and Ryu, J.
\newblock Randomized algorithms accelerated over cpu-gpu for ultra-high
  dimensional similarity search.
\newblock In \emph{ACM SIGMOD Record}, pp.\  889--903. ACM, 2018.

\bibitem[Wicaksono et~al.(2011)Wicaksono, Tolubaeva, and Chapman]{besar2011fs}
Wicaksono, B., Tolubaeva, M., and Chapman, B.
\newblock Detecting false sharing in openmp applications using the darwin
  framework.
\newblock In \emph{Lecture Notes in Computer Science}, pp.\  282--288, 2011.

\bibitem[Yagnik et~al.(2011)Yagnik, Strelow, Ross, and Lin]{yagnik2011power}
Yagnik, J., Strelow, D., Ross, D.~A., and Lin, R.-s.
\newblock The power of comparative reasoning.
\newblock In \emph{2011 International Conference on Computer Vision}, pp.\
  2431--2438. IEEE, 2011.

\bibitem[Yen et~al.(2018)Yen, Kale, Yu, Holtmann-Rice, Kumar, and
  Ravikumar]{yen2018loss}
Yen, I. E.-H., Kale, S., Yu, F., Holtmann-Rice, D., Kumar, S., and Ravikumar,
  P.
\newblock Loss decomposition for fast learning in large output spaces.
\newblock In \emph{International Conference on Machine Learning}, pp.\
  5626--5635, 2018.

\end{thebibliography}
\bibliographystyle{mlsys2020}

%%%%%%%%%%%%%%%%%%%%%%%%%%%%%%%%%%%%%%%%%%%%%%%%%%%%%%%%%%%%%%%%%%%%%%%%%%%%%%%
%%%%%%%%%%%%%%%%%%%%%%%%%%%%%%%%%%%%%%%%%%%%%%%%%%%%%%%%%%%%%%%%%%%%%%%%%%%%%%%
% SUPPLEMENTAL CONTENT AS APPENDIX AFTER REFERENCES
%%%%%%%%%%%%%%%%%%%%%%%%%%%%%%%%%%%%%%%%%%%%%%%%%%%%%%%%%%%%%%%%%%%%%%%%%%%%%%%
%%%%%%%%%%%%%%%%%%%%%%%%%%%%%%%%%%%%%%%%%%%%%%%%%%%%%%%%%%%%%%%%%%%%%%%%%%%%%%%
\appendix

\section{Different Hash Functions}
\label{sec:hashfunctions}
\textbf{Signed Random Projection (Simhash) : } Refer \citep{gionis1999similarity} for explanation of the theory behind Simhash. We use $K \times L$ number of random pre-generated vectors with components taking only three values  $\{+1, 0, -1\}$. The reason behind using only $+1s$ and $-1s$ is for fast implementation. It requires additions rather than multiplications, thereby reducing the computation and speeding up the hashing process. 
To further optimize the cost of Simhash in practice, we can adopt the sparse random projection idea \citep{li2006very}. %The basic idea is to shrink the dimension of the random vectors. However, this creates a mismatch between the dimension of the input and the random vectors while computing the inner product. 
A simple implementation is to treat the random vectors as sparse vectors and store their nonzero indices in addition to the signs. For instance,  let the input vector for Simhash be in $R^d$. Suppose we want to maintain $1/3$ sparsity, we may uniformly generate $K*L$ set of $d/3$ indices from $[0, d-1]$. In this way, the number of multiplications for one inner product operation during the generation of the hash codes would simply reduce from $d$ to $d/3$. Since the random indices are produced from one-time generation, the cost can be safely ignored.

\textbf{Winner Takes All Hashing (WTA hash) : } In SLIDE, we slightly modify the WTA hash algorithm from \citep{yagnik2011power} for memory optimization. Originally, WTA takes $O(KLd)$ space to store the random permutations $\Theta$ given the input vector is in $R^d$. $m<<d$ is a adjustable hyper-parameter. We only generate $\frac{KLm}{d}$ rather than $K*L$ permutations and thereby reducing the space to $O(KLm)$. Every permutation is split into $\frac{d}{m}$ parts (bins) evenly and each of them can be used to generate one WTA hash code. Computing the WTA hash codes also takes $O(KLm)$ operations.

\textbf{Densified Winner Takes All Hashing (DWTA hash) : } As argued in \citep{chen2018densified}, when the input vector is very sparse, WTA hashing no longer produces representative hash codes. Therefore, we use DWTA hashing, the solution proposed in \citep{chen2018densified}. Similar to WTA hash, we generate $\frac{KLm}{d}$ number of permutations and every permutation is split into $\frac{d}{m}$ bins. DWTA loops through all the nonzero (NNZ) indices of the sparse input. For each of them, we update the current maximum index of the corresponding bins according to the mapping in each permutation. 

It should be noted that the number of comparisons and memory lookups in this step is $O(NNZ*\frac{KLm}{d})$, which is significantly more efficient than simply applying WTA hash to sparse input. For empty bins, the densification scheme proposed in \citep{chen2018densified} is applied.

\textbf{Densified One Permutation Minwise Hashing(DOPH): } The implementation mostly follows the description of DOPH in \citep{shrivastava2014densifying}. DOPH is mainly designed for binary inputs. However, the weights of the inputs for each layer are unlikely to be binary. We use a thresholding heuristic for transforming the input vector to binary representation before applying DOPH. The $k$ highest values among all $d$ dimensions of the input vector are converted to $1$s and the rest of them become $0$s. Define $idx_{k}$ as the indices of the top $k$ values for input vector $x$. Formally, 
\vspace{-2mm}
$$\text{Threshold}(x_i)=\begin{cases}
1, & \text{if $i \in idx_{k}$}.\\
0, & \text{otherwise}.
\end{cases}$$
We could use sorting algorithms to get the top $k$ indices, but it induces at least $O(dlogd)$ overhead. Therefore, we keep a priority queue with indices as keys and the corresponding data values as values. This requires $O(dlogk)$ operations. 

\section{Reducing the Sampling Overhead}\label{sec:sampling2}
The key idea of using LSH for adaptive sampling of neurons with large activation is sketched in `Introduction to overall system' section in the main paper. We have designed three strategies to sample large inner products: 1) Vanilla Sampling 2) Topk Sampling 3) Hard Thresholding. We first introduce them one after the other and then discuss their utility and efficiency. Further experiments are reported in section \ref{sec:design_choices}.

{\bf Vanilla Sampling: } Denote $\beta_l$ as the number of active neurons we target to retrieve in layer $l$. After computing the hash codes of the input, we randomly choose a table and only retrieve the neurons in that table. We continue retrieving neurons from another random table until $\beta_l$ neurons are selected or all the tables have been looked up. Let us assume we retrieve from $\tau$ tables in total. Formally, the probability that a neuron $N_l^j$ gets chosen is,
\begin{equation}
Pr(N_l^j) = (p^K)^{\tau}(1-p^K)^{L-\tau},
\end{equation}
where $p$ is the collision probability of the LSH function that SLIDE uses. For instance, if Simhash is used, $$p = 1 - \frac{cos^{-1}\left( \frac{(w_l^j)^Tx_l}{||w_l^j||_2\cdot ||x_l||_2}\right)}{\pi}.$$ 
From the previous process, we can see that the time complexity of vanilla sampling is $O(\beta_l)$.

{\bf TopK Sampling: } In this strategy, the basic idea is to obtain those neurons that occur more frequently among all $L$ hash tables. After querying with the input, we first retrieve all the neurons from the corresponding bucket in each hash table. While retrieving, we use a hashmap to keep track of the frequency with which each neuron appears. The hashmap is sorted based on the frequencies, and only the neurons with top $\beta_l$ frequencies are selected. This requires additional $O(\vert N_l^a \vert)$ space for maintaining the hashmap and $O(\vert N_l^a \vert+ \vert N_l^a \vert log \vert N_l^a \vert)$ time for both sampling and sorting. 

\begin{figure}[tb]
	\centering
	\mbox{
		\includegraphics[width=3in]{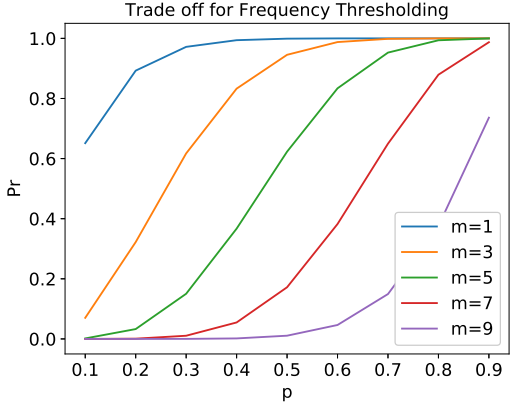}
	}
	\caption{Hard Thresholding: Theoretical selection probability $Pr$ vs the collision probabilities $p$ for various values of frequency threshold $m$ (eqn. \ref{eqn:pr_vs_p}). High threshold ($m=9$) gets less number of false positive neurons but misses out on many active neurons. A low threshold ($m=1$) would select most of the active neurons along with lot of false positives.}
	\label{fig:thres}
	\vspace{-3mm}
\end{figure}

{\bf Hard Thresholding:} The TopK Sampling could be expensive due to the sorting step. To overcome this, we propose a simple variant that collects all neurons that occur more than a certain frequency. This bypasses the sorting step and also provides a guarantee on the quality of sampled neurons. Suppose we only select neurons that appear at least $m$ times in the retrieved buckets, the probability that a neuron $N_l^j$ gets chosen is,
\vspace{-2mm}
\begin{equation}\label{eqn:pr_vs_p}
    Pr(N_l^j) = \sum_{i = m}^{L} \tbinom {L}{i}(p^K)^{i}(1-p^K)^{L-i},
\end{equation}
Figure \ref{fig:thres} shows a sweep of curves that present the relation between collision probability of $h_l(w_l^j)$ and $h_l(x_l)$  and the probability that neuron $N_l^j$ is selected under various values of $m$ when $L=10$. We can visualize the trade-off between collecting more good neurons and omitting bad neurons by tweaking $m$. For a high threshold like $m=9$, only the neurons with $p>0.8$ have more than $Pr>0.5$ chance of retrieval. This ensures that bad neurons are eliminated but the retrieved set might be insufficient. However, for a low threshold like $m=1$, all good neurons are collected but bad neurons with $p<0.2$ are also collected with $Pr>0.8$. Therefore, depending on the tolerance for bad neurons, we choose an intermediate $m$ in practice.  

\begin{figure}[t]
% 	\label{softmax}
	\centering
	\mbox{
		\includegraphics[width=3in]{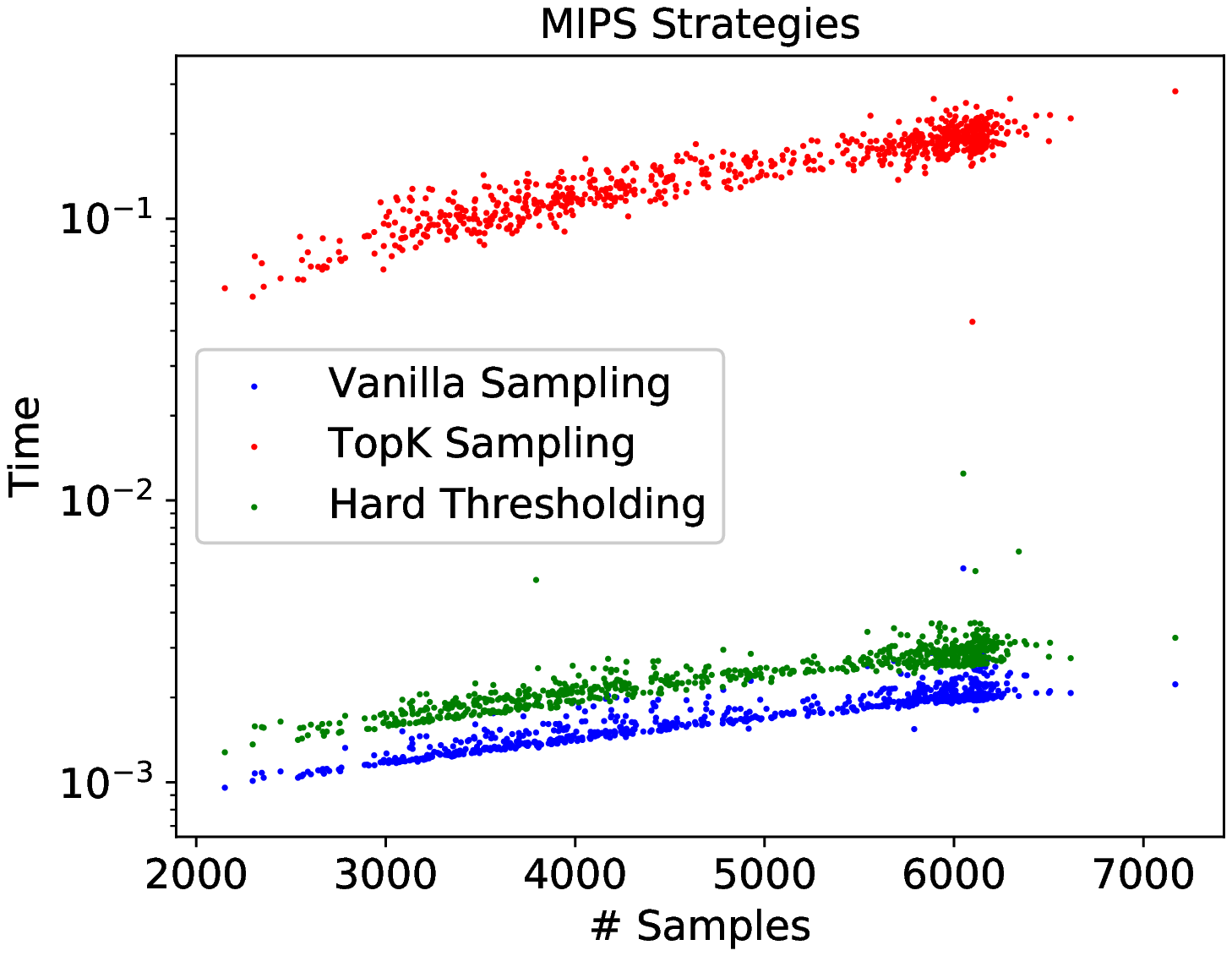}
	}
	\caption{Sampling Strategies: Time consumed (in seconds) for various sampling methods after retrieving active neurons from Hash Tables.}
	\label{fig:sort2}
% 	\vspace{-0.5cm}
\vspace{-3mm}
\end{figure}

\section{Design Choice Comparisons}\label{sec:design_choices}
In the main paper, we presented several design choices in SLIDE which have different trade-offs and performance behavior, e.g., executing MIPS efficiently to select active neurons, adopting the optimal policies for neurons insertion in hash tables, etc. In this section, we substantiate those design choices with key metrics and insights. In order to better analyze them in more practical settings, we choose to benchmark them in real classification tasks on Delicious-200K dataset.

\subsection{Evaluating Sampling Strategies}
Sampling is a crucial step in SLIDE. The quality and quantity of selected neurons and the overhead of the selection strategy significantly affect the SLIDE performance. We profile the running time of these strategies, including Vanilla sampling, TopK thresholding, and Hard thresholding, for selecting a different number of neurons from the hash tables during the first epoch of the classification task.

Figure \ref{fig:sort2} presents the results. The blue, red and green dots represent Vanilla sampling, TopK thresholding, and Hard thresholding respectively. It shows that the TopK thresholding strategy takes magnitudes more time than Vanilla sampling and Hard thresholding across all number of samples consistently. Also, we can see that the green dots are just slightly higher than the blue dots meaning that the time complexity of Hard Thresholding is slightly higher than Vanilla Sampling. Note that the y-axis is in log scale. Therefore when the number of samples increases, the rates of change for the red dots are much more than those of the others. This is not surprising because TopK thresholding strategy is based on sorting algorithms which has $O(nlogn)$ running time. Therefore, in practice, we suggest choosing either of Vanilla Sampling or Hard Thresholding for efficiency. For instance, we use Vanilla Sampling in our extreme classification experiments because it is the most efficient one. Furthermore, the difference between iteration wise convergence of the tasks with TopK Thresholding and Vanilla Sampling are negligible.

\subsection{Addition to Hashtables}
SLIDE supports two implementations of insertion policies for hash tables described in section~\ref{subsec:design} in main paper. We profile the running time of the two strategies, Reservoir Sampling and FIFO. After the weights and hash tables initialization, we clock the time of both strategies for insertions of all 205,443 neurons in the last layer of the network, where 205,443 is the number of classes for Delicious dataset. Then we also benchmark the time of whole insertion process including generating the hash codes for each neuron before inserting them into hash tables.

The results are shown in Table \ref{fig:table}. The column ``Full Insertion" represents the overall time for the process of adding all neurons to hash tables. The column ``Insertion to HT" represents the exact time of adding all the neurons to hash tables excluding the time for computing the hash codes. Reservoir Sampling strategy is more efficient than FIFO. From an algorithmic view, Reservoir Sampling inserts based on some probability, but FIFO guarantees successful insertions. We observe that there are more memory accesses with FIFO. However, compared to the full insertion time, the benefits of Reservoir Sampling are still negligible. Therefore we can choose either strategy based on practical utility. For instance, we use FIFO in our experiments.

\begin{table}
	\label{fig:table}
	\centering
		\caption{Time taken by hash table insertion schemes}
	\begin{tabular}{|c|c|l|} \hline
		& Insertion to HT & Full Insertion\\ \hline
		Reservoir Sampling & 0.371 s    & 18 s\\ \hline
		FIFO               & 0.762 s    & 18 s  \\
		\hline\end{tabular}
\end{table}

\section{Threading Model and Platform Micro-architecture Optimization}\label{sec:HPC}

Our experimental analysis shows that SLIDE is a memory-bound workload. We show that a careful workload optimization to design a threading model and a data access pattern to take into consideration the underlying platform architecture leads to a significant performance boost. 

% With the introduction of Intel's Skylake processoers~\citep{skylake}, new micro-architecture features are available; for example, support for AVX-512 instructions with two 512-bit wide floating-point multiply accumulate. Increase in compute throughput was supported by doubling of L1-D cache bandwidth and quadrupling of unified L2 cache capacity. However, the shared and distributed last-level on-chip cache (LLC) was changed from an all-inclusive model that holds a copy of any data item in any core private caches to a non-inclusive LLC to make more effective use of on-chip cache capacity. Hence, private cores' data are not necessarily included in the shared cache among cores, and as a result, without a thoughtful threading model design, threads can frequently stall, especially in memory bound workloads, due to misses for shared data that are held in cores' private caches. Next, we provide all the thoughtful threading optimizations that we incorporate into SLIDE.    

\textbf{OpenMP and Cache Optimizations:} A key metric for the identification of memory and cache performance bottlenecks in a multi-threaded application, e.g., SLIDE, is the number of data misses in the core private caches. This is a significant source of coherence traffic, potentially making the shared bus a bottleneck in a symmetric multiprocessor (SMP) architecture, thus increasing memory latency.

OpenMP provides a standard, easy to use model for scaling up a workload among all the available platform cores. The master thread forks a specified number of worker threads to run concurrently, and by default, threads are kept unbound and are spread across available cores, if any. Generally speaking, an inclusive last level cache (LLC) can improve data sharing, because a new thread created to run on a remote core, can probably find a copy of a shared data structure in the LLC, this is especially true if the accesses are mostly read-only, and ignoring the effect of evictions overhead from private core caches~\citep{meng09llc}. With the new trend in CPU architecture of a non-inclusive LLC (e.g. Intel's Skylake architecture~\cite{skylake}) multi-threaded workloads can operate on larger data per thread (due to increased L2 size). However, due to the new design of a non-inlusive LLC remote thread missing on a shared data structure can cause cache thrashing, invalidation, and bouncing of shared data among cores. We noticed that SLIDE is prone to this bottleneck.
 
Fortunately, OpenMP provides a control for thread affinity where a mask is set by an affinity preference and checked during runtime for possible locations for a thread to run. When threads are accessing mostly private independent data items, it is best to scatter these among the available possible cores for an almost linear speedup with the available cores due to no data dependency. On the other hand, if these threads are accessing items in a shared data structure, it is generally better to schedule these threads in a more compact packing (using the OpenMP Affinity=close) where threads are scheduled closer (same CPU socket) as the master thread. 

Furthermore, CPU caches are arranged into cache lines. Multiple threads updating data items that happen to co-locate into the same cache line (called false sharing) can also cause cache thrashing, since these updates need to be serialized to ensure correctness, leading to performance degradation. Much previous work (e.g., ~\citep{besar2011fs}) have tried to detect and resolve the issue of false sharing for OpenMP multi-threads mainly using compiler optimizations and hardware performance counters. However, generally speaking, carefully allocating data structures and aligning them on cache line boundaries (e.g., by padding) significantly reduce the false sharing opportunities.  We chose to use the later alternative for SLIDE.  

\textbf{Address Translation and Support for Kernel Hugepages:}\label{sec:cache_thrashing} Virtual memory provides applications with a flat address space and an illusion of sufficiently large and linear memory. The addressed memory is divided into fixed-size pages, and a page table is used to map virtual pages to physical ones. The address lookup is accelerated using Translation Lookaside Buffers (TLBs).

Since SLIDE is a workload with a large memory footprint, the performance of virtual memory paging can suffer due to stagnant TLB sizes. TLB address translation is on the processors’ critical path. It requires low access times which constrain TLB size (and thus, the number of pages it holds). On a TLB miss, the system must walk the page table, which may incur additional cache misses. Recent studies show that workloads with large memory footprints can experience a significant performance overhead due to excessive page table walks~\citep{karakos2014paging,basu2013paging}.

We employ Hugepages for SLIDE, which is a technology for x86-64 architectures to map much larger pages than the default 4KB normal-sized pages on the orders of 2 MB to 1 GB. Use of huge pages (Transparent Hugepages and libhugetlbfs~\citep{Corbet2011THP}) increases TLB reach substantially, and reduces the overhead associated with excessive TLB misses and table walks.

\textbf{Vector Processing, Software Pipelining, and Prefetching:} 
We further use software optimization techniques to improve workload performance in SLIDE. In particular, we use Vector processing which is capable of exploiting data-level parallelism through the use of Single-Instruction-Multiple-Data (SIMD) execution, where a function is called with a batch of inputs instead of an individual input (e.g., the function to update a large matrix of weights in the back-propagation phase). The implementation uses SIMD instructions (e.g., Intel AVX~\citep{skylake}) to implement the update to multiple weights simultaneously. Implementing a software pipeline is an excellent way to hide memory latency for memory-bound workloads. Our implementation divides the processing of data items into stages of a pipeline, where explicit software prefetch stage (using, for example, x86 PREFETCHT0 instruction set) is followed by a processing stage(s). The data items that are accessed in the future are prefetched into the core caches in advance to the time when they are needed to get processed. In particular, for a vector processing of updating of N weights, a software implementation can prefetch weight $W_{i+d}$ (where d is the depth of the pipeline) while updating weight $W_i$, as a result, when it is time to process weight $W_{i+d}$ it is already in the CPU cache. 

\subsection{Measuring the Impact of Transparent Hugepages}\label{sec:tlb}
In table~\ref{tab:tlb}, we show the results for examining the impact of Transparent Hugepages on various CPU-counter metrics.

A direct benefit of employing Transparent Hugepages is the drastic reduction in TLB miss rate. For example, the first row in table \ref{tab:tlb} shows that the TLB load miss rate for data reduces from $5.12\%$ to $0.25\%$. Similarly, TLB load miss rate for instruction also decreases from $56.12\%$ to $20.96\%$. Consequently, we expect a huge reduction in page table walks (PTW) incurred due to TLB misses. This is corroborated in rows 3 and 4 of table \ref{tab:tlb}. We see that the ratios of CPU cycles spent by PTWs caused by data and instruction TLB misses have reduced from $7.74\%$ to $0.72\%$ and $0.02\%$ to $0.015\%$ respectively. As mentioned in section \ref{sec:cache_thrashing}, TLB misses cause expensive main memory reads. Using Hugepages, we reduce the memory reads caused by data and instruction TLB misses from $3,062,039/s$ to $749,485/s$ and $12,060/s$ and $11,580/s$ respectively. Finally, we also report the reduction in page faults (which can possibly occur when there is a TLB miss) from $32,548/s$ to $26,527/s$.

\begin{table}[h]
  \centering
  \begin{tabular}{p{3.5cm}p{1.7cm}p{1.7cm}} %lllll
    \toprule
    Metric & Without Hugepages & With Hugepages \\
    \midrule
    dTLB load miss rate &  $5.12\%$ & $0.25\%$\\
    iTLB load miss rate &  $56.12\%$ & $20.96\%$\\
    PTW dTLB-miss & $7.74\%$  & $0.72\%$\\
    PTW iTLB-miss & $0.02\%$ & $0.015\%$\\
    RAM read dTLB-miss & $3,062,039/s$ & $749,485/s$ \\
    RAM read iTLB-miss & $12,060/s$ & $11,580/s$\\
    PageFault & $32,548/s$  & $26,527/s$ \\
    \bottomrule
  \end{tabular}
  \caption{Comparison of various CPU-counter metrics for both cases; with and without using Transparent Hugepages.}
  \label{tab:tlb}
\end{table}      

\section{More discussion on scalability}
\begin{figure}[h]
	\centering
	\mbox{
		\includegraphics[width=1.7in]{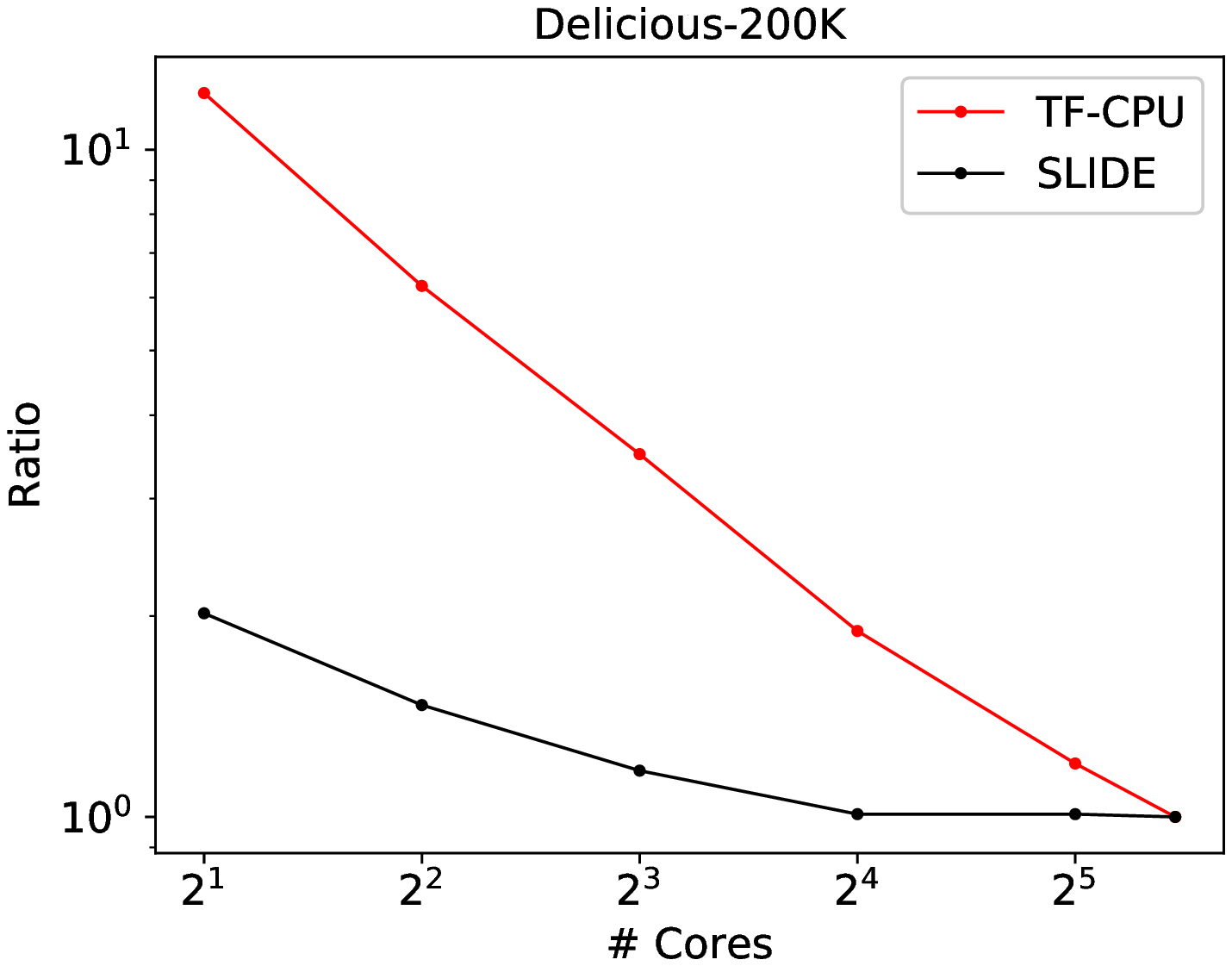}
        \includegraphics[width=1.7in]{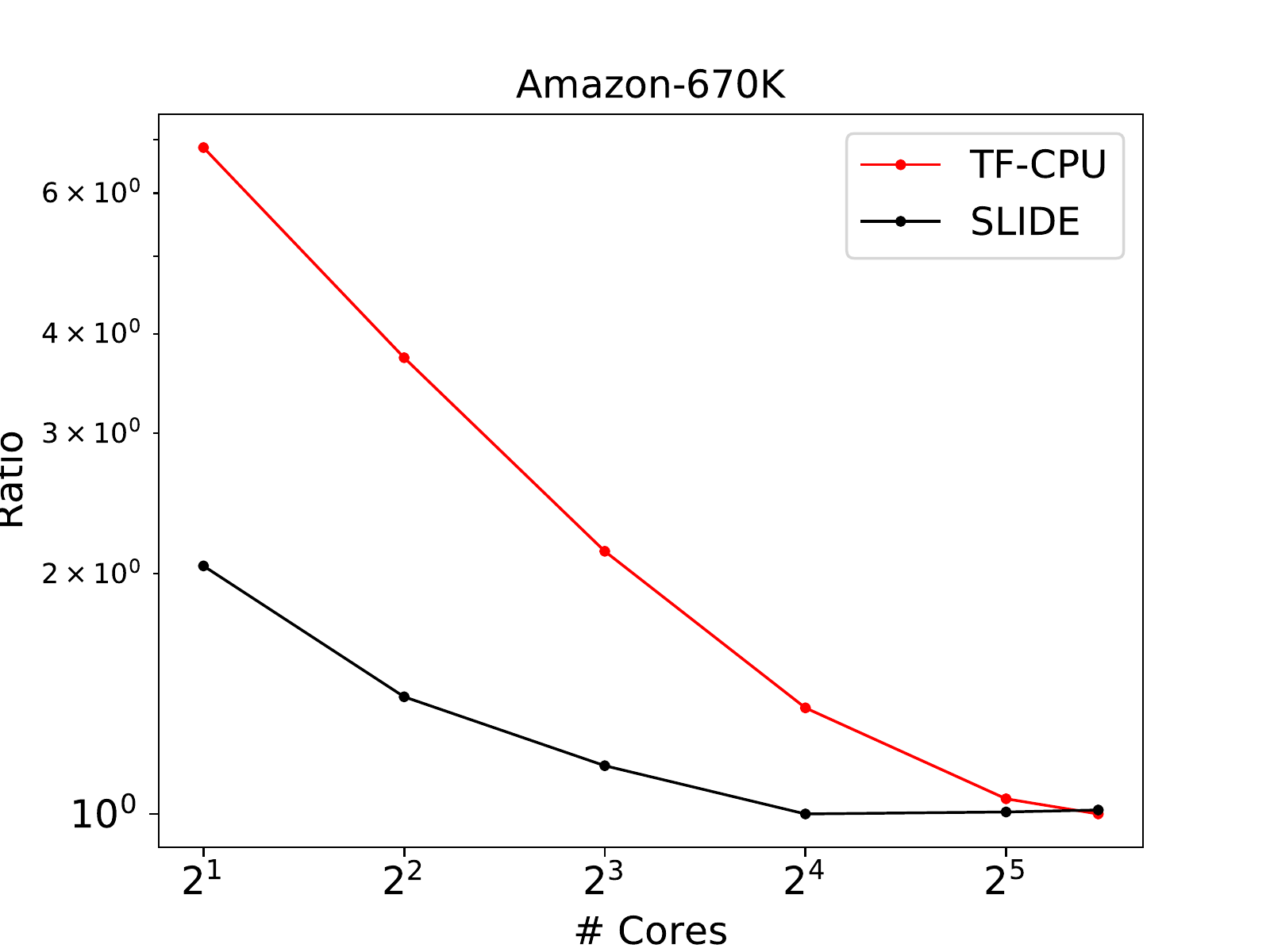}
	}
	\caption{Scalability Tests: Comparison of performance gains with the number of CPU cores for SLIDE (in red ) vs. Tensorflow-CPU (in black) vs. Tensorflow-GPU (in blue). The blue line is flat because the performance of TF-GPU does not depend on CPU cores. We notice that the convergence time drops steeply for SLIDE compared to TF-CPU/GPU. On Delicious-200K dataset, SLIDE beats TF-CPU with just 8 cores and TF-GPU with less than 32 cores. Similarly, on Amazon-670K dataset, SLIDE beats TF-CPU with just 2 cores and TF-GPU with just 8 cores. The $2^{nd}$ and $4^{th}$ plots compare the ratio of the convergence time at a various number of CPU cores to the minimum time required (when we use all 44 CPU cores).}
	\label{fig:scale2}
\end{figure}

Moreover, based on the statistics collected through experiments as mentioned above, we show the ratio of convergence time with the different number of cores to the minimum convergence time (using 44 cores). The results are exhibited in Figure \ref{fig:scale2}. Again, the red line represents SLIDE, and the black line represents Tensorflow-CPU. When the number of cores increases, that ratio decreases for both SLIDE and Tensorflow-CPU. However, it is explicit that the ratio drops more drastically for the red line than the black line. This behavior concludes that the scalability of SLIDE is much better than that of Tensorflow-CPU. Moreover, in the plot, we observe that the benefits of using more cores are not obvious after 16 cores for Tensorflow-CPU. Coincidentally, a very recent work \citep{DBLP:journals/corr/abs-1812-01665} introduces the hardness of finding the optimal parameter settings of Tensorflow’s threading model for CPU backends. It argues that getting the best performance from a CPU needs manual, tedious and time-consuming tuning and it still may not guarantee the best performance. While analyzing the scalability and core utilization of Tensorflow-CPU can be an independent research interest, we explore a small aspect of it in the following paragraphs.

%%%%%%%%%%%%%%%%%%%%%%%%%%%%%%%%%%%%%%%%%%%%%%%%%%%%%%%%%%%%%%%%%%%%%%%%%%%%%%%
%%%%%%%%%%%%%%%%%%%%%%%%%%%%%%%%%%%%%%%%%%%%%%%%%%%%%%%%%%%%%%%%%%%%%%%%%%%%%%%

\end{document}